\documentclass[aps,prb,twocolumn,showpacs,10pt]{revtex4-1}
\pdfoutput=1
\usepackage{epsfig}
\usepackage{amsmath}
\usepackage{amssymb}
\usepackage[colorlinks=true,citecolor=blue,linkcolor=blue]{hyperref}
\newcommand{\bs}{\boldsymbol}

\begin{document}

\title{Fractionally charged topological point defects on the kagome lattice}
\author{Andreas R\"uegg}
\author{Gregory A. Fiete}
\affiliation{Department of Physics, The University of Texas at Austin, Austin, Texas 78712, USA}
\date{\today}

%%%%%%%%
\begin{abstract}
We study a two-dimensional system of spin-polarized fermions on the kagome lattice at filling fraction $f=1/3$ interacting through a nearest-neighbor interaction $V$. Above a critical interaction strength $V_c$ a charge-density wave with a broken $Z_3$ symmetry is stabilized. Using the unrestricted mean-field approximation, we present several arguments showing that elementary topological point defects in the order parameter bind a fractional charge. Our analysis makes use of two appealing properties of the model: (i) For weak interaction, the low-energy degrees of freedom are described by Dirac fermions coupled to a complex-valued mass field (order parameter). (ii) The nearest-neighbor interaction is geometrically frustrated at filling $f=1/3$. Both properties offer a route to fractionalization and yield a consistent value $\pm1/2$ for the fractional charge as long as the symmetry between the up and the down triangles of the kagome lattice is preserved. If this symmetry is violated, the value of the bound charge varies continuously with the strength of the symmetry-breaking term in the model. In addition, we have numerically computed the confining potential between two fractionally charged defects. We find that it grows linearly at large distances but can show a minimum at a finite separation for intermediate interactions. This indicates that the polaron state, formed upon doping the charge-density wave, can be viewed as a bound state of two defects.
\end{abstract}
%%%%%%%

\maketitle

%%%%%%%%%%%
\section{Introduction}
%%%%%%%%%%%
The concept of fermion fractionalization has been applied to a variety of condensed matter systems. Prominent examples are spin-charge separation in polyacetylene,\cite{Su:1979} fractionally charged excitations in the fractional quantum Hall states\cite{Laughlin:1999} and magnetic monopoles in spin-ice.\cite{Castelnovo:2008} In all these examples, excitations carrying fractional quantum numbers with respect to the elementary particles forming the system were found. The term fractionalization is stringently used only if well-defined excitations with fractional quantum number exist on all length scales. In gapped insulating systems in dimensions $d\geq 2$ this requires topological order of the ground state,\cite{Oshikawa:2006} i.e.~a ground state degeneracy which depends on the topology of the underlying system. Naturally, the effective low-energy theory describing the fractional excitations is a gauge theory in the deconfining phase, such as the Chern-Simons theory for the two dimensional fractional quantum Hall state or the Coulomb gauge theory for the three dimensional spin-ice materials and other frustrated magnetic systems.\cite{Balents:2002,Hermele:2004, Banerjee:2008} Electron fractionalization has also been discussed in the context of the high-$T_c$ cuprates on the basis of the $Z_2$ gauge theory.\cite{Senthil:2000} In this article, however, we would like to use the concept of fractionalization in a less stringent way. Particularly, we are interested in phenomena where fractionalization occurs only up to a certain length scale which usually depends on temperatures and model parameters. This more general point of view allows one to cover a broader range of phenomena and to access this fascinating phenomena from distinct theoretical view-points.

Several authors have stressed the field-theoretical point of view where fractionalized quantum numbers are carried by solitonic solutions of field theories which support isolated mid-gap states.\cite{Jackiw:1976, Goldstone:1981} In condensed matter physics, this route to fractionalization is well appreciated for the one-dimensional example discussed by Su, Schrieffer and Heeger where the soliton describes a domain wall separating two degenerate dimerized ground states.\cite{Su:1979} More recently, a body of work has appeared\cite{Mudry:2007,Jackiw:2007,Chamon:2008, Chamon:2008b,Seradjeh:2008b, Guo:2009, Hou:2010,Liu:2010,Weeks:2010,Wang:2010} on generalizations of this concept to two dimensions and it has been argued that topological defects in the ``kekule" order parameter offers an example for the solitonic fractionalization in two-dimensional graphene.\cite{Mudry:2007} However, in contrast to the one-dimensional case, the energy cost associated with a vortex in the complex-valued Bose field is not finite but grows with system size. In the continuum limit, the interaction between two vortices depends logarithmically on the distance and vortices can proliferate above the Kosterlitz-Thouless temperature. On the lattice, however, they are always confined at sufficiently long distances.\cite{Mudry:2007,Seradjeh:2008b} It was also pointed out that in order to heal the vortex at long distances, a coupling to an axial gauge field (which itself supports a vortex) can be introduced, thereby rendering the vortex energy finite.\cite{Jackiw:2007,Chamon:2008b}

Another perspective on the fractionalization phenomena emerges from models describing strongly interacting particles on a lattice with geometrical frustration.\cite{Fulde:2002,Runge:2004} In this class of models, the strong interaction enforces a local constraint and it is the violation of this local constraint which carries a fractional charge. Clearly, there is a close relation to frustrated spin models and the afore mentioned spin-ice system is a prominent representative. In many cases, the frustrated particle interaction or spin exchange can be mapped on an effective hardcore dimer model. Removing one dimer introduces two monomers which, under certain circumstances, are well defined fractionalized excitations.\cite{Rokhsar:1988,Moessner:2001a,Fradkin:2004,Sikora:2009} 

In this article, we want to make contact to both routes to fractionalization by studying topological point defects in a model of spin-polarized fermions on the kagome lattice subject to a nearest-neighbor interaction $V$. Previous investigations\cite{Wen:2010,Nishimoto:2010} of this model at filling fraction $f=1/3$ suggest a zero-temperature phase transition at a critical interaction $V_c$ between the semi-metallic Dirac liquid for $V<V_c$ and a gapped and charge ordered state with a $\sqrt{3}\times\sqrt{3}$ reconstruction of the unit cell for $V>V_c$. Our discussion of topological point defects in the order parameter of the charge-density wave will make use of two important properties of the model. First, the low-energy degrees of freedom in the weakly interacting limit are well-described by Dirac-fermions coupled to the complex-valued order parameter which enters as a mass field. This offers the possibility for the solitonic fractionalization mechanism in two-dimensions in analogy with graphene\cite{Mudry:2007, Jackiw:2007,Chamon:2008, Chamon:2008b,Liu:2010} and related systems.\cite{Seradjeh:2008b,Weeks:2010,Wang:2010} Second, the nearest-neighbor interaction is geometrically frustrated and the classical charge configurations with lowest energy satisfy the ``triangle rule". This constraint states that there is exactly one particle on every triangle of the kagome lattice. A local violation binds a fractional charge. Indeed, this possibility has recently been explored in the strongly interacting limit using exact diagonalization techniques and it has been argued that the defects carry a fractional charge $\pm1/2$ and are asymptotically free in the large $V$ limit.\cite{OBrien:2010}

Our paper is organized as follows. In Sec.~\ref{sec:model} we introduce the model and discuss the triangle rule and the effect of its local violation. In Sec.~\ref{sec:meanfield} we introduce the unrestricted mean-field approximation, discuss the leading instability of the Dirac liquid towards the charge-density wave and introduce a Ginzburg-Landau expansion of the free energy. This sets the stage for introducing topological point defects in Sec.~\ref{sec:vortices} and in Sec.~\ref{sec:numerics} we numerically study solutions with point defects, compute the value of the bound charge and the confinement potential between two defects. Eventually, in Sec.~\ref{sec:Dirac}, we consider the weakly ordered state and establish a description in terms of Dirac fermions coupled to a complex-valued mass term.

%%%%%%%%%%%%%%%%%%%%%%%%%%%
\section{Model for charge-ordered kagome lattice}
\label{sec:model}
%%%%%%%%%%%%%%%%%%%%%%%%%%%
Our starting point is a tight-binding model of spin-polarized fermions on the kagome lattice at filling fraction $f=1/3$ subject to a nearest-neighbor repulsion $V$. The Hamiltonian is given by
\begin{equation}
H=-t\sum_{\langle i,j\rangle}\left( c_i^{\dag}c_j^{}+{\rm h.c.}\right)+V\sum_{\langle i,j\rangle}n_in_j+H_{\rm BOW}.
\label{eq:model}
\end{equation}
Here, $c_i^{(\dag)}$ annihilates (creates) a spin-polarized fermion on site $i$ and $n_i=c_i^{\dag}c_i^{}$. The hopping integral is denoted by $t>0$ and $V>0$ specifies the nearest-neighbor interaction. 
At several places in this article, we will also consider a term which enhances the hopping on the up triangles
\begin{equation}
H_{\rm BOW}=-\delta t\sum_{\langle i,j\rangle\in\Delta}\left(c_i^{\dag}c_j+{\rm h.c.}\right).
\label{eq:HBOW}
\end{equation}
This term induces a bond-order wave which breaks the symmetry between the up and the down triangles. However, unless otherwise stated, we set $\delta t=0$. The non-interacting ($V=0$) band structure is obtained by diagonalizing the matrix $H_0({\bs K})=-2t\Gamma({\bs K})$ where
\begin{equation}
\Gamma({\bs K})=
\begin{bmatrix}
0 &\cos(K_1/2)&\cos(K_2/2)\\
\cos(K_1/2) & 0&\cos(K_3/2)\\
\cos(K_2/2)& \cos(K_3/2) &0
\end{bmatrix}.
\label{eq:Gamma}
\end{equation}
Above, we have introduced $K_{\nu}={\bs K}\cdot{\bs a}_{\nu}$ and ${\bs a}_{\nu}$ ($\nu=1,2,3$) are given in Tab.~\ref{tab:def}. There is a flat band at energy $2t$ as well as two dispersing bands. It is well-known that at $f=1/3$, the linearized band structure near the Fermi energy is described by two Dirac cones, similar to the situation found in graphene.
%%%%%%% 
\begin{figure}
\centering
\includegraphics[width=1\linewidth]{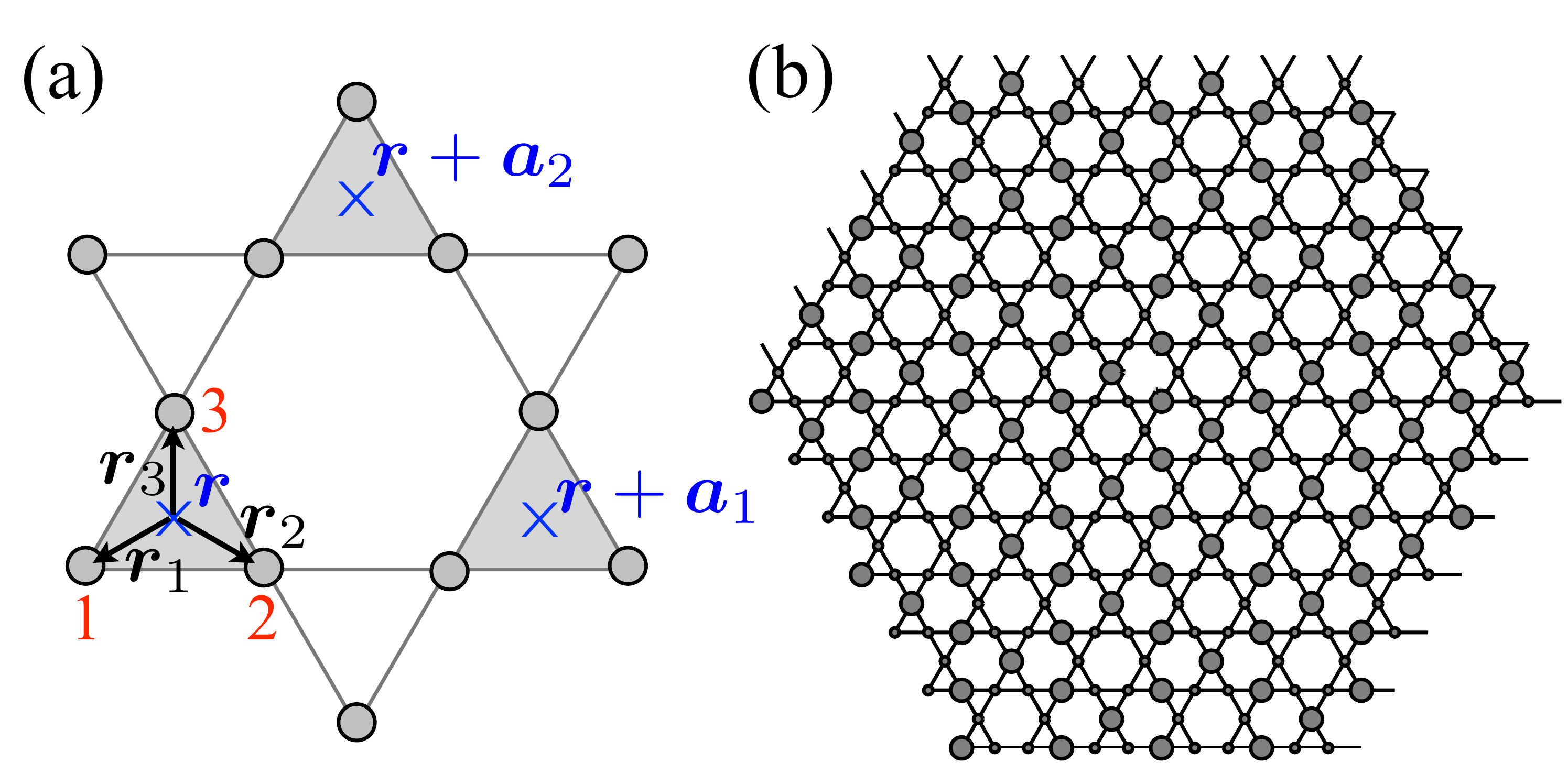}
\caption{(a) The unit cell of the kagome lattice contains three sites labeled by 1,2 and 3. The vector ${\bs r}$ is an element of the underlying triangle lattice and denotes the location of the center of the up-triangles. The unit cell vectors are denoted by ${\bs a}_1$ and ${\bs a}_2$. (b) Density distribution in the charge density wave phase found in the model defined by Eq.~\eqref{eq:model}. There is a $\sqrt{3}\times\sqrt{3}$ reconstruction of the unit cell.}
\label{fig:intro}
\end{figure}
%%%%%%

%%%%%%
\subsection{Triangle rule in the atomic limit}
%%%%%%
Let us now look at the atomic limit $t=0$. It is known that at filling fraction $1/3$ the interaction energy can be minimized by a macroscopic number of classical charge configurations. This fact becomes clear when rewriting the interaction Hamiltonian in {\it real space} as a sum over {\it all} triangles $\delta$ of the kagome lattice in the following way:
\begin{equation}
H_V=V\sum_{\langle i,j\rangle}n_in_j=\frac{V}{2}\sum_{\delta}(N_{\delta}-1)^2,
\end{equation}
where $N_{\delta}=\sum_{i\in \delta}n_i$ denotes the total charge operator on the triangle $\delta$ ($\delta$ can label both an up or a down triangle). Clearly, the interaction is lowest for configurations which fulfill the local constraint $N_{\delta}=1$. Taking ${\bs r}$ to be the center of an up-triangle and using the labeling convention introduced in Fig.~\ref{fig:intro}(a) and Tab.~\ref{tab:def} this constraint takes the form
\begin{subequations}
\label{eq:icerule}
\begin{align}
n_1({\bs r})+n_2({\bs r})+n_3({\bs r})&=1,\label{eq:a}\\
n_3({\bs r})+n_1({\bs r}+{\bs a}_2)+n_2({\bs r}+{\bs a}_3)&=1.\label{eq:b}
\end{align}
\end{subequations}
The first equation is written for the up and the second one for the down triangles. Thus, $H_V$ is minimized by classical charge configurations with exactly one fermion on every triangle and the ground state is macroscopically degenerate. In the following, we refer to the local constraint Eq.~\eqref{eq:icerule} as the ``triangle rule".\cite{Bergman:2006}
%%%%%%%
\begin{table}[ht]
\center
\begin{tabular}{r||c|c|c}
$\nu$&1&2&3\\
\hline
\hline
${\bs a}_{\nu}$&(1,0)&$(1/2,\sqrt{3}/2)$&$(-1/2,\sqrt{3}/2)$\\
${\bs G}\cdot{\bs a}_{\nu}$&$2\pi/3$&$-2\pi/3$&$2\pi/3$\\
\hline
${\bs r}_{\nu}$&$(-1/4,-\sqrt{3}/12)$&$(1/4,-\sqrt{3}/12)$&$(0,\sqrt{3}/6)$\\
${\bs G}\cdot{\bs r}_{\nu}$&$2\pi/3$&$-2\pi/3$&$0$
\end{tabular}
\caption{Definitions of the lattice vectors used in this paper (in units of the lattice constant $a$) and the values of the inner product with the uniform ordering vector ${\bs G}=(8\pi/3,0)$ modulo $2\pi$.}
\label{tab:def}
\end{table}
%%%%%%
It is instructive to see how the macroscopic degeneracy shows up in {\it reciprocal space}. Introducing the Fourier components
\begin{equation}
n_{\nu}({\bs Q})=\sum_{\bs r}n_{\nu}({\bs r})e^{-i{\bs Q}\cdot({\bs r}+{\bs r}_{\nu})}
\end{equation}
we can write the interaction as
\begin{equation}
H_V=\frac{V}{N}\sum_{\bs Q}\vec{n}({\bs Q})^{\dag}\Gamma({\bs Q})\vec{n}({\bs Q}).
\end{equation}
Here, $N$ denotes the number of unit cells and we have introduced the vector notation
\begin{equation}
\vec{n}({\bs Q})^{\dag}=
\begin{bmatrix}
n_1(-{\bs Q})&n_2(-{\bs Q})&n_3(-{\bs Q})
\end{bmatrix}.
\end{equation}
The matrix $\Gamma({\bs Q})$ is given in Eq.~\eqref{eq:Gamma}. Its lowest eigenvalue is equal to $-1$, independent of ${\bs Q}$. It follows that the interaction energy is minimized by all charge configurations which have Fourier components lying in the flat band. Indeed, it is straightforward to show that charge configurations which are proportional to the eigenvectors of the flat band fulfill the two constraints Eq.~\eqref{eq:icerule} in momentum space. For ${\bs Q}=0$ the conditions \eqref{eq:a} and \eqref{eq:b} are equivalent which is a manifestation of the quadratic band touching point at ${\bs Q}=0$ in $\Gamma({\bs Q})$.

The macroscopic degeneracy of the classical charge configurations is lifted for finite $t$. In particular, for $t/V\ll 1$, the model Eq.~\eqref{eq:model} can be mapped onto a quantum dimer model by identifying an occupied site of the kagome lattice with a dimer on the hexagonal lattice.\cite{Nishimoto:2010,OBrien:2010} Thereby, ring exchange processes of order $t^3/V^2$ in the original model translate into dimer flips in the dimer model which stabilizes a valence bond crystal with a $\sqrt{3}\times\sqrt{3}$ reconstructed unit cell.\cite{Moessner:2001} The kinetic energy gained by resonating plaquettes favors charge configurations which are connected by local dimer flips. The classical configuration which has most flippable plaquettes corresponds to the mean-field charge-density wave shown in Fig.~\ref{fig:intro}(b). Note also that the constraint \eqref{eq:icerule} maps onto a hardcore constraint for dimer coverings. 
%%%%%
\subsection{Violation of triangle rule and fractional charge}
\label{sec:violation}
\begin{figure}
\centering
\includegraphics[width=0.6\linewidth]{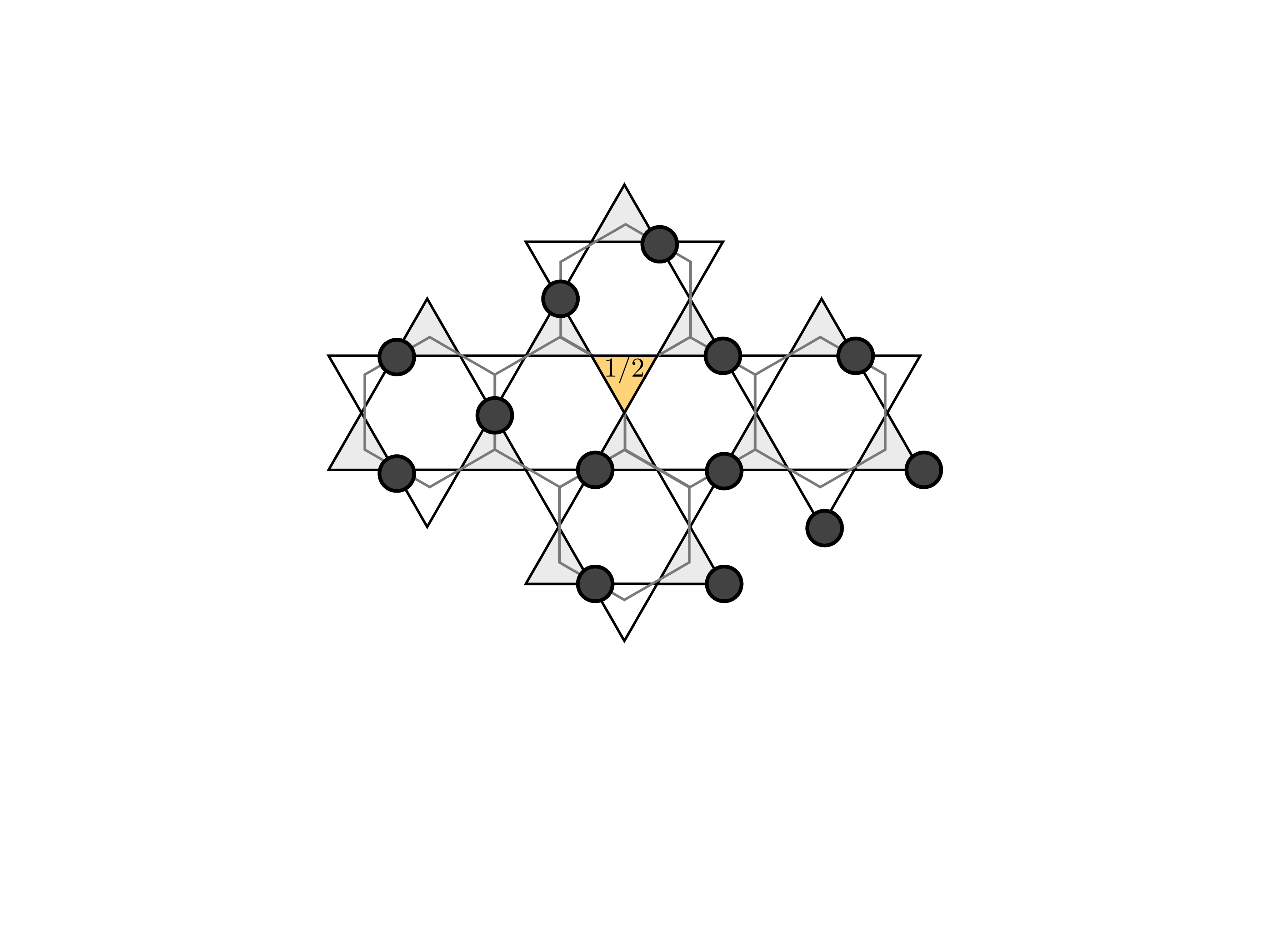}
\caption{(Color online.) A classical charge configuration which locally violates the triangle rule on the shaded (yellow) down triangle. We introduce a local charge density defined on every triangle as half the value of the sum of the charges on that triangle. In this way we see that the violation of the triangle rule carries a fractional charge $1/2$.}
\label{fig:violation}
\end{figure}

%%%%%
Let us now consider a classical charge configuration which locally violates the triangle rule Eq.~\eqref{eq:icerule} either for an up or a down triangle. An example is shown in Fig.~\ref{fig:violation} where the triangle rule is violated on a single down triangle. In such a situation, the total charge density per unit cell depends on how it is measured! For example, if we measured it by summing the charges on the up triangles, we would conclude that there is exactly one particle in every unit cell. On the other hand, if we measured it by summing the charges on the down triangles, we would conclude that there is one particle missing in the unit cell which contains the empty down triangle. A more sensitive way which avoids this ambiguity is to introduce a charge density defined on {\it every} triangle as half the value of the sum of the charges on that triangle. This charge density is then defined on the hexagonal lattice formed by the center points of the triangles, see Fig.~\ref{fig:violation}. If the triangle rule is fulfilled everywhere, there is a charge density $-1/2$ on every site of the hexagonal lattice (we associate a charge $-1$ with a single particle). In this way we see that an empty triangle carries a fractional charge $1/2$ compared to a configuration which satisfies the triangle rule. Likewise, a triangle with two particles carries a charge $-1/2$ and with three particles a charge $-1$.
%%%
\section{Mean-field description}
\label{sec:meanfield}
%%%
In this section we start with a conventional mean-field theory and discuss some properties of the uniformly ordered system. In the Hartree approximation the density-density interaction is decoupled in the following way:
\begin{eqnarray}
&&n_{\nu}({\bs r})n_{\mu}({\bs r}')\approx \nonumber\\
&&n_{\nu}({\bs r})\rho_{\mu}({\bs r}')+n_{\mu}({\bs r}')\rho_{\nu}({\bs r})-\rho_{\nu}({\bs r})\rho_{\mu}({\bs r}').
\label{eq:Hartree}
\end{eqnarray}
Here, we have introduced the expectation values 
\begin{equation}
\rho_{\nu}({\bs r})=\langle n_{\nu}({\bs r})\rangle=\frac{1}{N}\sum_{\bs Q}\rho_{\nu}({\bs Q})e^{i{\bs Q}\cdot({\bs r}+{\bs r}_{\nu})}
\end{equation}
of the local densities with Fourier components $\rho_{\nu}({\bs Q})$. In addition to the Hartree terms in Eq.~\eqref{eq:Hartree} also the Fock terms have been considered in Ref.~\onlinecite{Wen:2010} for uniform solutions. These terms tend to stabilize the semi-metallic phase and the critical interaction strength for the phase transition is $V_c^{\rm HF}\approx 3t$, in good agreement with other methods.\cite{Nishimoto:2010} On the other hand, when keeping only the Hartree terms as in Eq.~\eqref{eq:Hartree}, the critical interaction strength is smaller, $V_c^{\rm H}\approx 2.2t$. However, except for this shift, the qualitative aspects of the Hartree solution seems to be the same and to keep it simple, we use the decoupling Eq.~\eqref{eq:Hartree}. We note here that the situation for filling fraction $f=2/3$ is quite different because complex Fock terms stabilize an interaction-driven topological insulator for arbitrary small nearest-neighbor interactions,\cite{Wen:2010} similar to what is found on the decorated honeycomb lattice at half filling.\cite{Ruegg:2010}
\subsection{Mean-field triangle rule}
Any mean-field state is characterized by a self-consistent charge distribution $\{\rho_{\nu}({\bs r})\}$ and the configurations with lowest energies fulfill the triangle-rule Eq.~\eqref{eq:icerule} on average. In Fourier space, we can write it for ${\bs Q}\neq 0$ as
\begin{subequations}
\label{eq:mfconstraint}
\begin{align}
0&=\rho_{1}({\bs Q})e^{-iQ_2/2}+\rho_{2}({\bs Q})e^{-iQ_3/2}+\rho_{3}({\bs Q}),\\
0&=\rho_{1}({\bs Q})e^{iQ_2/2}+\rho_{2}({\bs Q})e^{iQ_3/2}+\rho_{3}({\bs Q}),
\end{align}
\end{subequations}
where $Q_{\nu}={\bs Q}\cdot{\bs a}_{\nu}$, as before. If we introduce the vector $\vec{\rho}({\bs Q})=[\rho_1({\bs Q})\, \rho_2({\bs Q})\, \rho_3({\bs Q})]^{T}$ the above condition is equivalent to the ``flat-band" condition
\begin{equation}
\Gamma({\bs Q})\vec{\rho}({\bs Q})=-\vec{\rho}({\bs Q}).
\label{eq:flatband}
\end{equation}
The mean-field interaction can then be written as
\begin{equation*}
H_V'=-\frac{2V}{N}\sum_{{\bs Q},\nu}n_{\nu}(-{\bs Q})\rho_{\nu}({\bs Q})+\frac{V}{N}\sum_{{\bs Q},\nu}\rho_{\nu}(-{\bs Q})\rho_{\nu}({\bs Q}),
\end{equation*}
where $\vec{\rho}({\bs Q})$ satisfies Eq.~\eqref{eq:flatband}.
\subsection{Leading instability}
\begin{figure}
\centering
\includegraphics[width=1\linewidth]{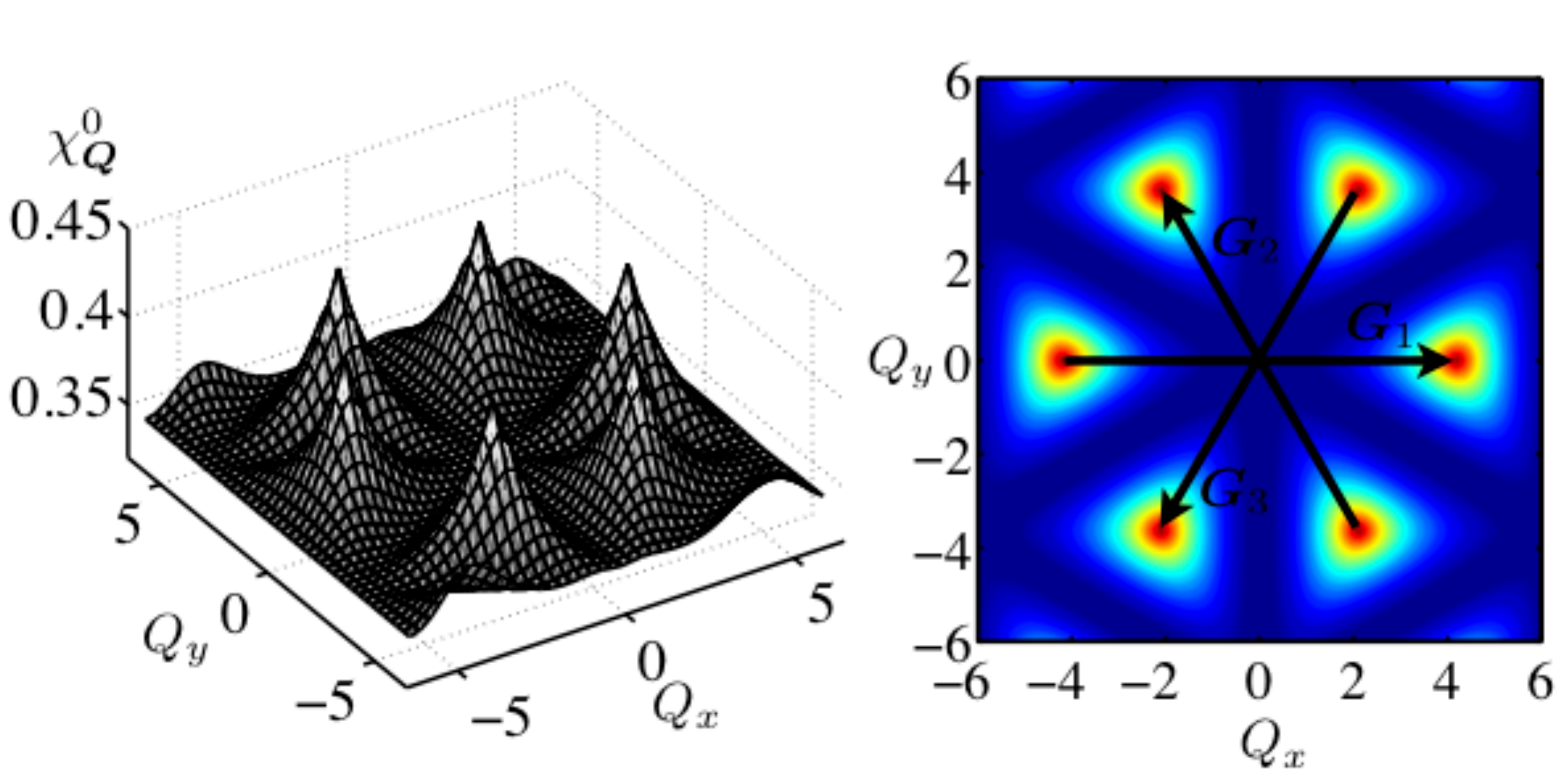}
\caption{(Color online.) The static susceptibility $\chi_{\bs Q}^0$ of the noninteracting system at $T=0$ associated with charge configurations satisfying the triangle rule on average. $\chi_{\bs Q}^0$ is largest at the corners of the hexagon forming the first Brillouin zone. The leading instability is a charge-density wave with one of the three ordering vectors ${\bs G}_1$, ${\bs G}_2$ or ${\bs G}_3$.}
\label{fig:chiQ0}
\end{figure}
In order to find the leading instability of the interacting system as function of the interaction $V$, we can study the static mean-field susceptibility associated with ``flat-band" configurations [configurations which are compatible with Eq.~\eqref{eq:flatband}]:
\begin{equation}
\chi_{\rm MF}({\bs Q})=\frac{\chi_{\bs Q}^0}{1-V\chi_{\bs Q}^0}.
\label{eq:chiMF}
\end{equation}
The leading instability at a fixed temperatures occurs at an ordering vector ${\bs G}$ which satisfies $V\chi_{\bs G}^0=1$ for the smallest value of $V$. This also defines the critical interaction $V_c^0=1/\chi_{\bs G}^0$. In Eq.~\eqref{eq:chiMF} we have introduced the response function of the non-interacting system which probes charge configurations satisfying the triangle rule on average:
\begin{equation*}
\chi^0_{\bs Q}(T)=-\frac{T}{N}\sum_{i\omega_n}{\rm Tr}\left[G_0(i\omega_n)A({\bs Q})G_0(i\omega_n)A(-{\bs Q})\right].
\end{equation*}
Here, $G_0(i\omega_n)=1/[i\omega_n-H_0+\mu]$ is the Matsubara Green's function operator of the non-interacting system and $\omega_n$ are fermionic Matsubara frequencies. The operator $A({\bs Q})=\vec{e}({\bs Q})\cdot \vec{n}({\bs Q})$ with $\Gamma({\bs Q})\vec{e}({\bs Q})=-\vec{e}({\bs Q})$ and $\vec{e}({\bs Q})\cdot \vec{e}({\bs Q})=1$ enforces the triangle rule on average. The trace involves summation over $\bs K$ and the three bands of the noninteracting system. Figure~\ref{fig:chiQ0} shows $\chi_{\bs Q}^0(T=0)$. The static susceptibility is largest at the corners of the hexagonal Brillouin zone and for the critical interaction at $T=0$ we find the numerical value $V_c^0(T=0)\approx 2.33t$. The leading instability therefore occurs at one of the three ordering vectors ${\bs G}_1$, ${\bs G}_2$ or ${\bs G}_3$ connecting opposite corners of the hexagon.

As mentioned earlier, this instability is a charge-density wave with a $\sqrt{3}\times\sqrt{3}$ reconstruction of the unit cell and is shown in Fig.~\ref{fig:intro}(b). Instead of working with three different ordering vectors, we fix ${\bs G}={\bs G}_1$ and allow for a complex phase of the charge-density wave order parameter, see below. In the following we choose
\begin{equation}
{\bs G}={\bs K}_+-{\bs K}_-
\end{equation}
where ${\bs K}_{\pm}$ denote the location of the two inequivalent Dirac points in the first Brillouin zone,
\begin{equation}
{\bs K}_{\pm}=\pm\left(\frac{4\pi}{3 a},0\right).
\label{eq:Kpm}
\end{equation}
Obviously, ${\bs G}$ couples the two Dirac points and it is this ``nesting" which opens a gap above a critical interaction.
%%%%%
\subsection{Ginzburg-Landau expansion}
%%%%%
From the triangle rule \eqref{eq:mfconstraint} it follows that $\rho_1({\bs G})=\rho_2({\bs G})=\rho_3({\bs G})$ and we define the complex-valued order parameter of the charge-density wave as
\begin{equation}
\Delta=|\Delta|e^{i\varphi}=-\frac{2V}{3N}[\rho_1({\bs G})+\rho_2({\bs G})+\rho_3({\bs G})].
\label{eq:Delta0}
\end{equation}
Note that there is an overall phase freedom in the definition of $\Delta$ and that the amplitude satisfies $|\Delta|\leq 2V/3$. With the definition Eq.~\eqref{eq:Delta0} of the order parameter, the interaction Hamiltonian for the uniform charge-density wave reduces to
%%%
\begin{equation}
H_V'=\Delta\sum_{\nu}n_{\nu}(-{\bs G})+\Delta^{*}\sum_{\nu}n_{\nu}({\bs G})+\frac{3N}{2V}|\Delta|^2.\\
\label{eq:HVDelta}
\end{equation}
%%%
The free energy of a slowly varying charge-density wave in the continuum limit is given by the perturbative Ginzburg-Landau expansion in terms of $\Delta$ and its gradient ${\vec \nabla}\Delta$:
\begin{eqnarray}
F_{\rm{CDW}}-F_0&=&\int\!\!\frac{dx\,dy}{\mathcal{A}}\,\, \Big[\alpha(V,T)|\Delta|^2+\eta|\Delta^*{\vec \nabla}\Delta|+\kappa|{\vec \nabla}\Delta|^2\nonumber\\
&&+ \gamma|\Delta|^3\cos(3\varphi)+\beta|\Delta|^4+\cdots\Big].
\label{eq:GL}
\end{eqnarray}
Here, $\mathcal{A}=\sqrt{3}a^2/2$ is the unit cell area of the kagome lattice and the coefficient $\beta>0$ stabilizes this expansion to order $|\Delta|^4$. The coefficient $\alpha(V,T)$ changes sign as function of the interaction strength or temperature and is given by
\begin{equation}
\alpha(V,T)=\frac{3}{2}\left[\frac{1}{V}-\frac{1}{V^0_c(T)}\right],\quad V_c^0(T)=1/\chi_{\bs G}^0(T).
\end{equation}
The term proportional to $\cos(3\varphi)$ in the expansion Eq.~\eqref{eq:GL} introduces an anisotropy as a result of the three-fold rotation symmetry of the triangular Bravais lattice. The numerical value of its prefactor at $T=0$ is $\gamma\approx 0.22/t^2$. This term acts as a pinning potential for the complex phase $\varphi$ of the order parameter $\Delta$ and in the ground state, it assumes one of the three values
\begin{equation}
\varphi_A=-\pi/3,\quad \varphi_B=\pi/3,\quad\varphi_C=\pi,
\label{eq:Z3}
\end{equation}
thereby reducing the continuous rotation symmetry to a three-fold one. This $Z_3$ freedom arises from the possibility to translate the configuration of the charge-density wave as a whole by a unit cell vector ${\bs a}_1$ or ${\bs a}_2$. A finite $\gamma$ also shifts the critical interaction strength $V_c$ to a smaller value compared to $V_c^0$. Moreover, it turns the second order (quantum) phase transition into a first-order one. A crystal-field term $\propto\cos(p\varphi)$ can also strongly affect the thermodynamic properties of the model and the value of the integer $p$ is important.\cite{Jose:1977} For the planar $xy$-model supplemented with a crystal-field term $\propto\cos(p\varphi)$ it has been shown that the ground state always has a broken symmetry. However, for $p\geq 4$ and at higher temperatures, there is a critical phase characterized by bound vortex-antivortex pairs similar to the one found in the absence of the crystal-field. They can unbind above the Kosterlitz-Thouless-Berezinskii temperature. For $p=3$, such a critical phase is absent in the planar model.
%%%%%%%%%%%%%%%%%%%%%%%%%%%
\section{$Z_3$-vortices}
\label{sec:vortices}
%%%%%%%%%%%%%%%%%%%%%%%%%%%

To study spatially fluctuating solutions we start again from the energy functional~\eqref{eq:GL}. We find that the gradient term proportional to $\eta$ appears in the expansion because of the Dirac-like single-particle spectrum in momentum space at $f=1/3$. However, this term disappears for slowly varying configurations once a gap is opened in the non-interacting system with finite $\delta t$. To keep our discussion simple we set $\eta=0$ in the following. In this case, there are two distinct length scales in the problem. One is the coherence length $\xi=\sqrt{\kappa/|\alpha|}$ which describes the characteristic length scale over which the amplitude of the order parameter changes. The second one is related to the anisotropy and naturally appears in the equation of motion for $\varphi$:
\begin{equation}
{\vec \nabla}^2\varphi=-\frac{1}{\lambda_p^2}\sin(p\,\varphi).
\label{eq:sG}
\end{equation}
This is the so-called sine-Gordon equation and in our model $p=3$. The characteristic length scale for the anisotropy is $\lambda_p=\sqrt{2\kappa/(p\gamma|\Delta|})$ and it is a sensible quantity as long as $|\Delta|\approx{\rm const}$. If the linear extension $L$ of the system satisfies $L\ll \lambda_p$ the anisotropy term on the right hand side can be neglected\cite{Hudak:1982} and we can consider a special class of (singular) solutions to ${\vec \nabla}^2\varphi=0$:
\begin{equation}
\varphi(x,y)=q\, {\rm Im}[\log(x+iy)].
\label{eq:phivortex}
\end{equation}
These vortex solutions have an integer nonzero topological charge (vorticity)
\begin{equation}
q=\frac{1}{2\pi}\oint_{C}{\vec \nabla}\varphi \cdot{\bs ds},
\end{equation}
where $C$ is a loop encircling the singularity at the origin. The energy of a single vortex configuration grows logarithmically with system size $L$. If $\xi\ll L\ll \lambda_p$ we expect that thermally excited vortex-antivortex pairs are present.

On the other hand, on length scales $L\gg\lambda_p$, the right hand side of Eq.~\eqref{eq:sG} can no-longer be neglected. Then, the simplest nontrivial solution is a domain wall between two degenerate ground states. An example describing a kink which extends along the $x$ axis is
\begin{equation}
\varphi(y)=-\frac{\pi}{p}+\frac{4}{p}{\rm atan}\left(e^{-\sqrt{p}y/\lambda_p}\right)
\label{eq:kink}
\end{equation}
and from the energy functional Eq.~\eqref{eq:GL} it follows that there is a finite energy per length associated with the domain wall. There exist also single and multi-vortex solutions of Eq.~\eqref{eq:sG} which are obtained by deforming the vortex solutions of the Laplace equation.\cite{Hudak:1982,Borisov:1985,Gouvea:1997, Sinitsyn:2004} (For $p=4$ and $q=\pm1$ a particularly simple explicit expression is known.) The single vortex-like solutions have the property that for $|{\bs R}|\ll\lambda_p$ they reduce to expression \eqref{eq:phivortex} whereas for $|{\bs R}|\gg \lambda_p$ the domain walls between the degenerate ground states are resolved, as in Eq.~\eqref{eq:kink}. As a result, the energy of such a $Z_p$ vortex eventually grows linearly with system size.

%%%%%%%%
\begin{figure}[b]
\centering
\includegraphics[width=0.48\linewidth]{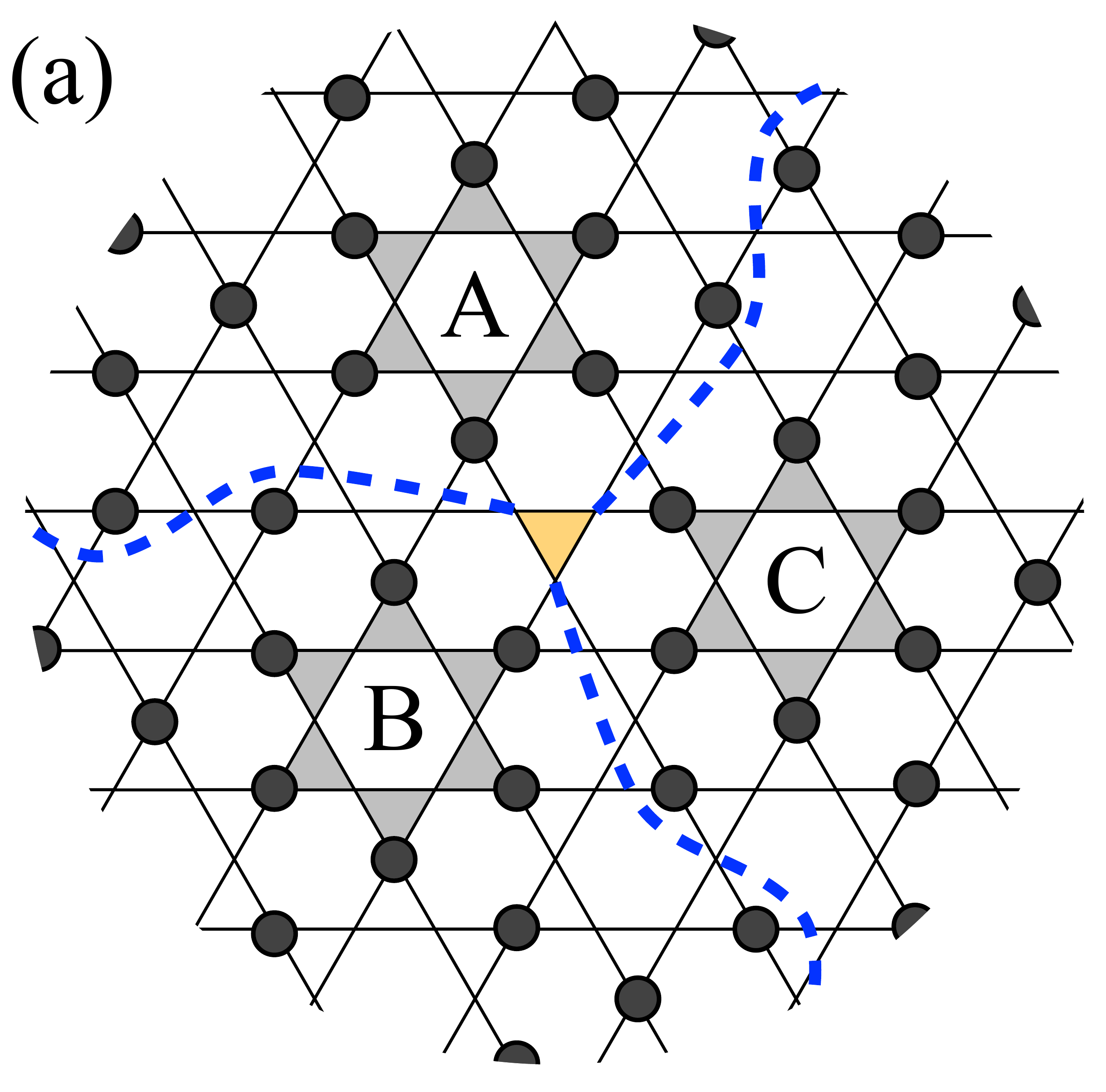}
\includegraphics[width=0.48\linewidth]{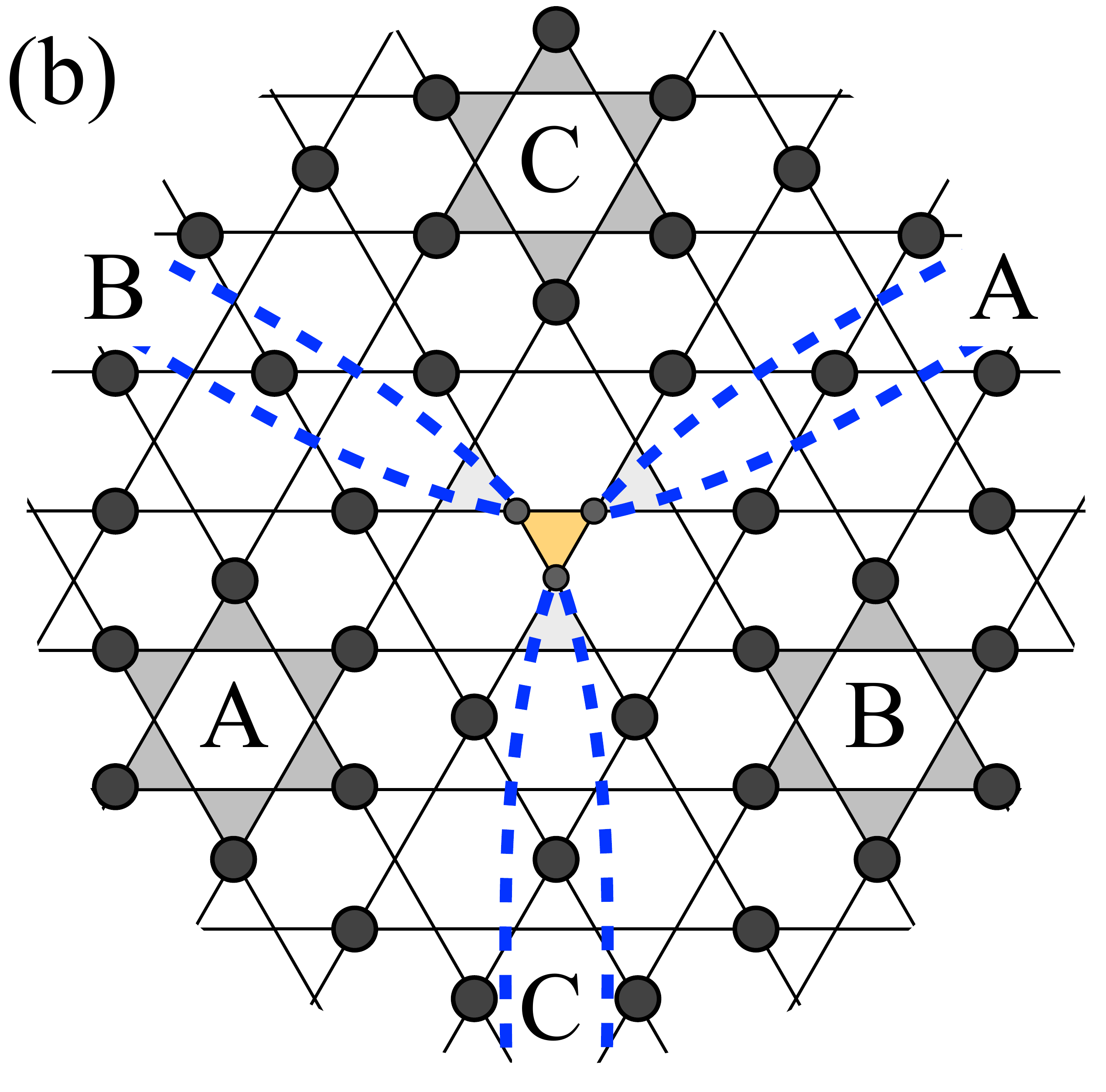}
\caption{(Color online.) Schematics of two different $Z_3$-vortices: A vortex with circulation $q=1$ is shown in (a) and a vortex with circulation $q=-2$ in (b). In (a), three domain walls meet at a down triangle and the triangle-rule is necessarily violated. As a consequence, a charge $1/2$ $(-1/2)$ is bound to this topological defect if the center triangle is empty (singly occupied). The configuration in (b) shows a double vortex where six domain walls meet. In the center there is one particle for one down triangle and three up triangles. This effectively results in two down triangle rule violations. The double vortex shown in (b) binds a charge 1 but it can be split into two topologically protected single vortices with a bound charge of 1/2.}
\label{fig:Z3}
\end{figure}
%%%%%%

%%%%%%%%%%%%%
\subsection{Triangle-rule violation in the vortex center}
%%%%%%%%%%%%%

%%%%%%
$Z_3$ vortices can also be considered in the classical limit and two examples are shown in Fig.~\ref{fig:Z3}. For a defect with $q=\pm1$, the complex phase of the order parameter changes from $\varphi_A$ to $\varphi_B$ to $\varphi_C$ and back to $\varphi_A$. This situations is sketched in Fig.~\ref{fig:Z3}(a). It turns out that for such an elementary defect there is necessarily one triangle (shaded) where the triangle rule is violated. For an up-triangle-rule violation (``up defect"), the phase changes clockwise while for a down-triangle-rule violation (``down defect") it changes counterclockwise. Furthermore, as explained in Sec.~\ref{sec:violation}, an empty triangle binds a (positive) deficit charge of $1/2$ compared to the uniform phase while a triangle with two fermions binds $-1/2$. Thus, we can label elementary defects by the pair $(Q,\delta)$ where $Q=\pm 1/2$ refers to the bound charge and $\delta=\triangle$ or $\triangledown$ indicates on which triangle the triangle rule is violated. These defects are topologically protected. Note that in the classical limit (vanishing hopping), domain walls do not cost any energy because the triangle rule is fulfilled.

It is also possible to construct defects which are composed of more then one elementary defect. An example is sketched in Fig.~\ref{fig:Z3}(b) where the phase changes twice when going clockwise around the defect. It can be viewed as a composite object of two elementary up defects $(1/2,\triangle)$ which in total binds a unit positive charge. Therefore, this defect is topologically not protected since it can be split into two elementary up defects. Another example is the composite object involving $(1/2,\triangle)$ and $(1/2,\triangledown)$ which can be viewed as a polaron state, see Sec.~\ref{sec:polaron}.

%%%%%%%

%%%%%%

%%%%%%%%%%%%%%%%%%%%%
\section{Numerical solutions with defects}
\label{sec:numerics}
%%%%%%%%%%%%%%%%%%%%%
We now turn to a numerical study of mean-field solutions with defects at zero temperatures. Thereby, we will focus on the properties of the elementary $Z_3$ vortices as sketched in Fig.~\ref{fig:Z3}(a). Because the defects are charged, we found it necessary to dope the system in order to stabilize mean-field solutions with defects. We therefore discuss examples where a single hole has been doped into a finite system with periodic boundary conditions. Self-consistent solutions are found by iterating the self-consistency equations. If the interaction is not too close to the critical interaction ($V\gtrsim 3t>V_c\approx 2.2t$), meaning that the defect size is comparable to the lattice constant, it is possible to choose the initial charge configuration such that solutions with two separated defects are stabilized.

%%%%%%%
\begin{figure}
\centering
\includegraphics[width=0.9\linewidth]{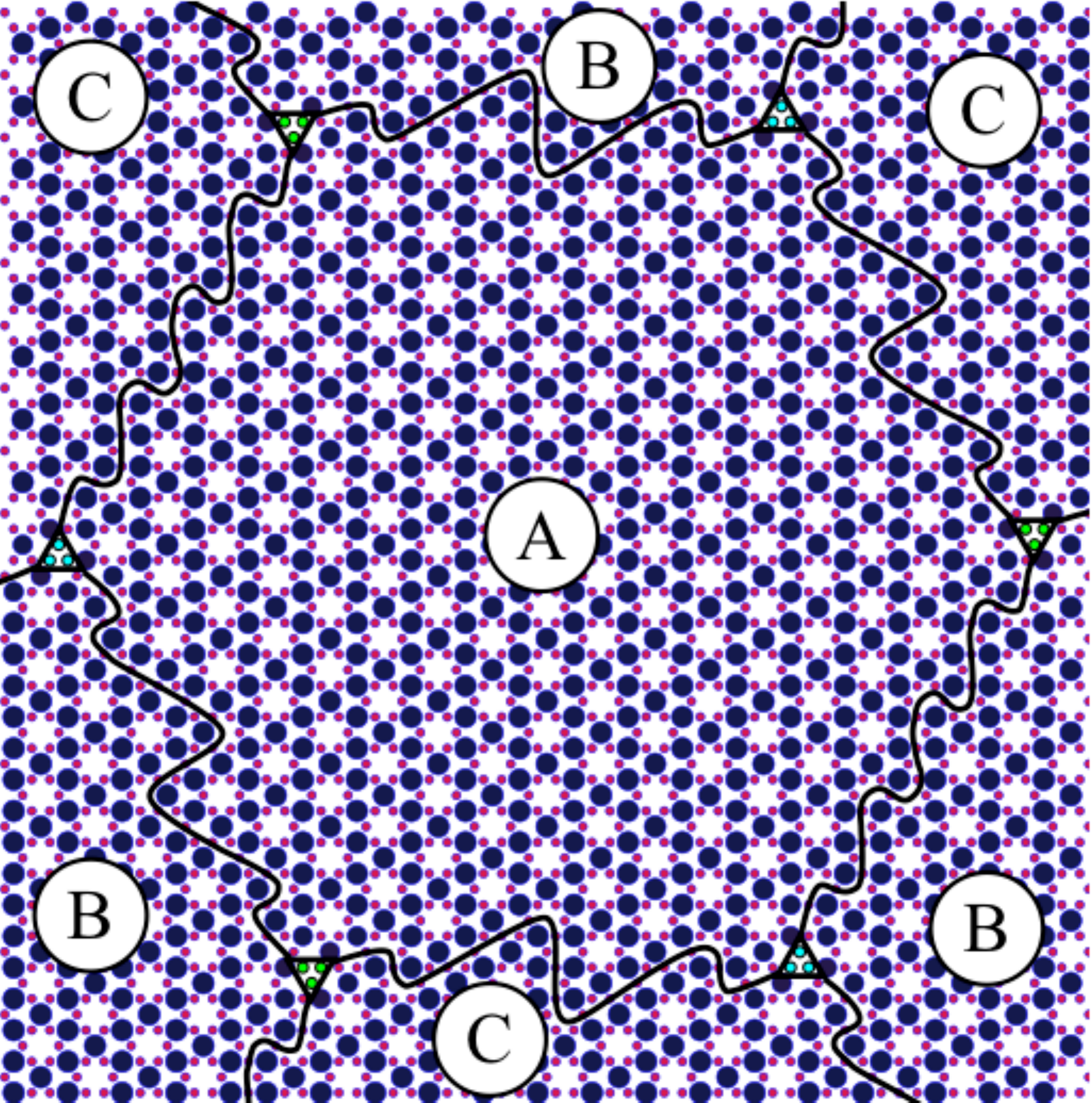}
\caption{(Color online.) Periodically extended charge configuration of a finite hexagonal system with defects at its corner. The considered system contains 1641 sites, 536 particles and two well-localized defects (up- and down triangles) each binding a charge 1/2. The three different uniform ground states (A, B and C) meet at the center of the defects and domain walls are indicated by the solid curves. The diameter of the circles building the kagome lattice is proportional to the local density and the interaction has been set to $V=4t$.}
\label{fig:configuration}
\end{figure}
%%%%%%

%%
\subsection{Fractionally charged defect}
Let us first look at a configuration where the defects form a regular lattice. In the most symmetric case, the up and down defects are arranged on interpenetrating triangular lattices and the defect lattice then has a hexagonal symmetry. Such a configuration offers a convenient possibility to investigate the properties of an isolated defect if they are separated far enough from each other. In actual calculations we considered finite systems of the form of a hexagon with defects located at its corners and employed periodic boundary conditions. Figure~\ref{fig:configuration} shows the self-consistent charge configuration (periodically extended) of such an arrangement. In this example, we have considered 1641 sites with 536 particles and the interaction has been set to $V=4t$. The diameter of the circles building the kagome lattice is proportional to the local density. As before, the three inequivalent uniform phases are denoted by A, B and C and in Fig.~\ref{fig:configuration} the domain walls between them are indicated as solid curves. In the initial state, the domain walls are straight lines but after the iteration process has converged they reveal a winding character.

\begin{figure}
\centering
\includegraphics[width=0.7\linewidth]{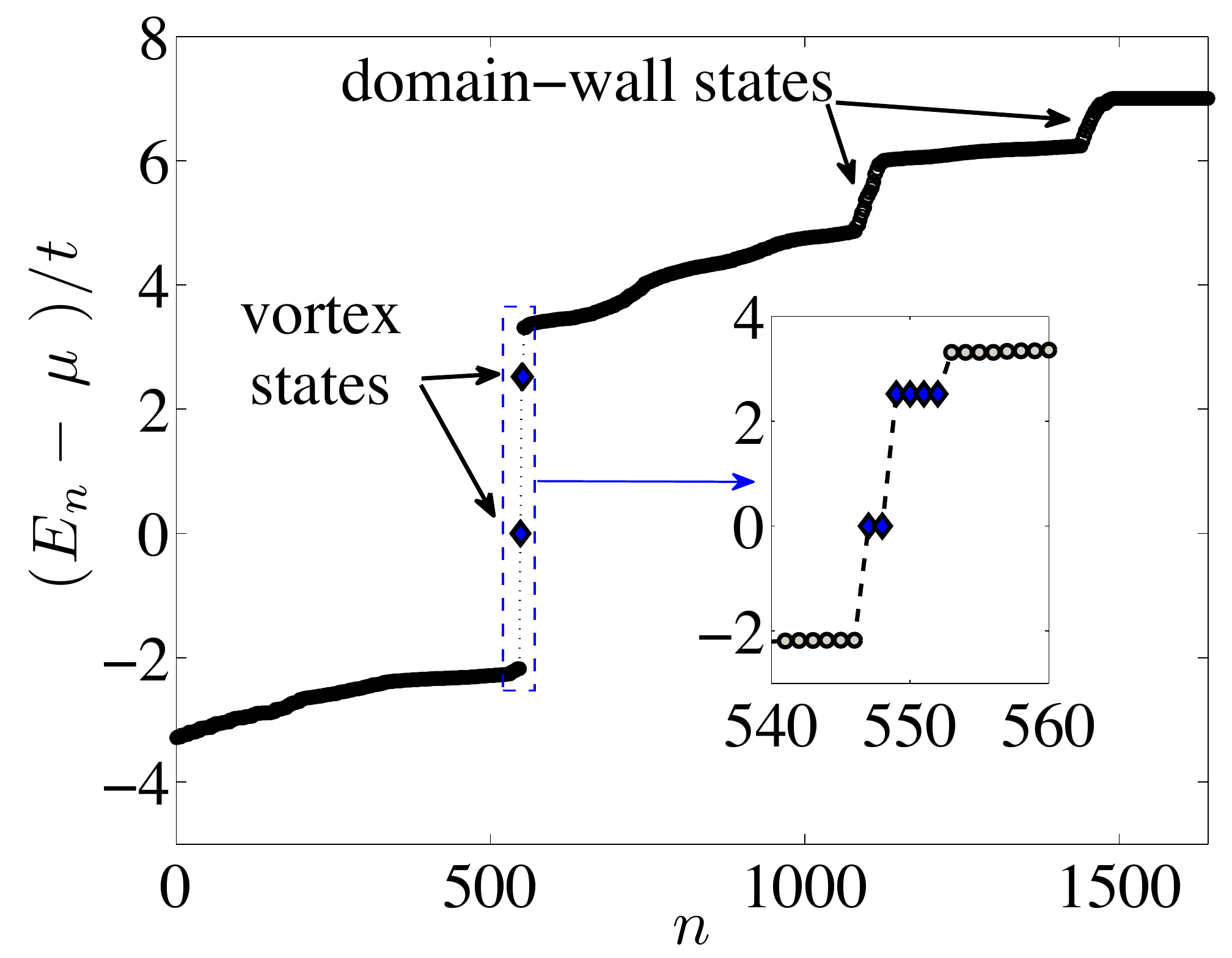}
\caption{Single-particle energy spectrum for the defect lattice shown in Fig.~\ref{fig:configuration}. The inset shows a blow-up of the spectrum near the gap. In total, there are six in-gap states which gives three per defect. The splitting between the upper four and lower two in-gap states is proportional to the hopping $t$ whereas the splitting among the upper or lower states is exponentially small. The quasi continuum of states between the step like features at higher energies are the domain-wall states.}
\label{fig:spectrum}
\end{figure}

Figure~\ref{fig:spectrum} shows the single particle energy spectrum of the same system. The defect states have energies which lie in the gap of the uniform phase. In total, there are six in-gap states which gives three states per defect. Out of the three states, one state is lower in energy than the other two. For the considered interaction strength, the defect size is comparable to the lattice spacing. This means that the defect states are basically localized on a single triangle and the energy splitting can be derived from the eigenenergies of a particle hopping on an isolated triangle (with amplitude $t$). Indeed, we have checked that in the large $V$ limit the energy splitting between the two upper states and the lower one approaches $3t=t-(-2t)$. Note that because the overlap between up and down-defects is very small for the considered system, the energy splitting between them is not visible in Fig.~\ref{fig:spectrum}. The step-like features in the spectrum at higher energies (near 6$t$ and $7t$) are the remains of the subband formation due to the enlarged unit cell of the uniform charge-density wave. For the uniform solution we find true energy gaps between the steps but in the presence of the topological defects, these gaps are filled. The states with energies between the steps are localized along the domain walls which can be confirmed by studying the wave-functions in real space.
%%%%%%%
%%%%%%

\begin{figure}
\centering
\includegraphics[width=1\linewidth]{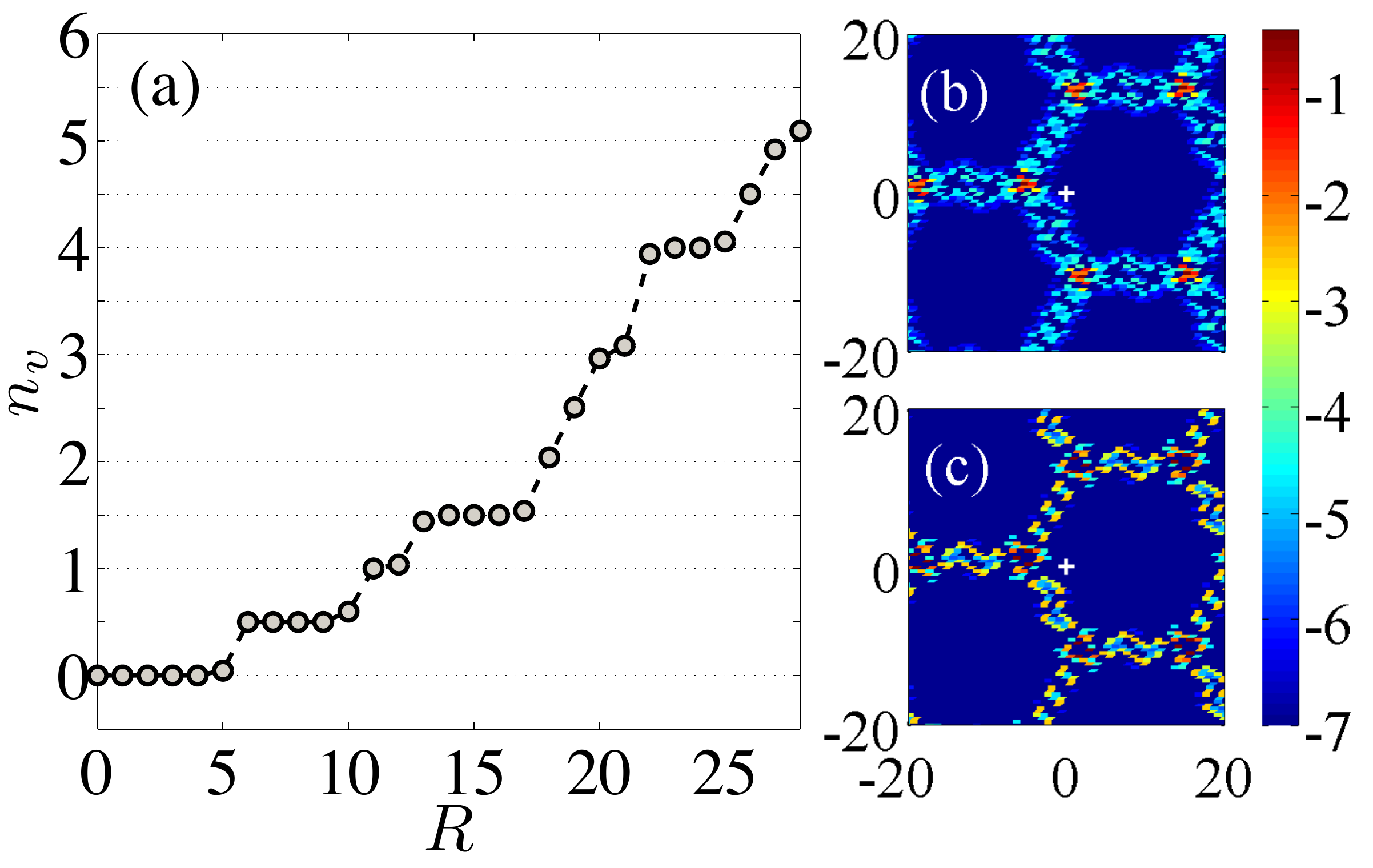}
\caption{Color online. (a) Integrated charge deficit within a circle of radius $R$ around the origin. The steps of magnitude $1/2$ indicate that a charge of $1/2$ is bound to each defect. (b) Logarithmic color plot of the locally averaged charge deficit and (c) logarithmic color plot of the locally averaged charge excess. In (b) and (c) a cut-off of $10^{-7}$ has been used; consistent with the numerical precision of the solution. The parameters of the defect lattice are the same as in Figs.~\ref{fig:configuration} and \ref{fig:spectrum}.}
\label{fig:density}
\end{figure}

From the analysis of classical charge configurations we expect that an elementary defect binds a fractional charge $\pm 1/2$. We now want to confirm this result for finite $t$ by an explicit calculation for the defect lattice of Fig.~\ref{fig:configuration}. In order to get rid of the short wave-length density oscillations, we have considered an averaged charge distribution on the hexagonal lattice defined by the center of mass points of the triangles forming the kagome lattice, as explained in Sec.~\ref{sec:violation}. Figure~\ref{fig:density}(a) shows the integrated density deficit (measured from the uniform particle density 1/2) within a circle of radius $R$ around the origin. The location of the origin has been chosen away from a high symmetry point of the defect lattice and is indicated in panel (b) and (c). As function of $R$, there are clearly visible steps of $1/2$ in the integrated deficit density which shows that every defect binds $1/2$ of charge. The steps are rather sharp indicating that the defects are rather well localized. This is also seen in panel (b) and (c) where the charge deficit and excess measured from $1/2$ in a logarithmic scale is shown. Indeed, most of the charge deficit is located very close to the defects but also along the domain walls, the density deviates from its uniform value. As a matter of fact, the charge density shows oscillatory behavior which is reminiscent of Friedel oscillations and the density can also exceed its uniform value in certain regions, see panel (c).
%%%%%%%
%%%%%%

%%%%%%%%%%%%%%%%%%%%%%%%%%
\subsection{Irrational versus rational charges}
\begin{figure}
\centering
\includegraphics[width=0.8\linewidth]{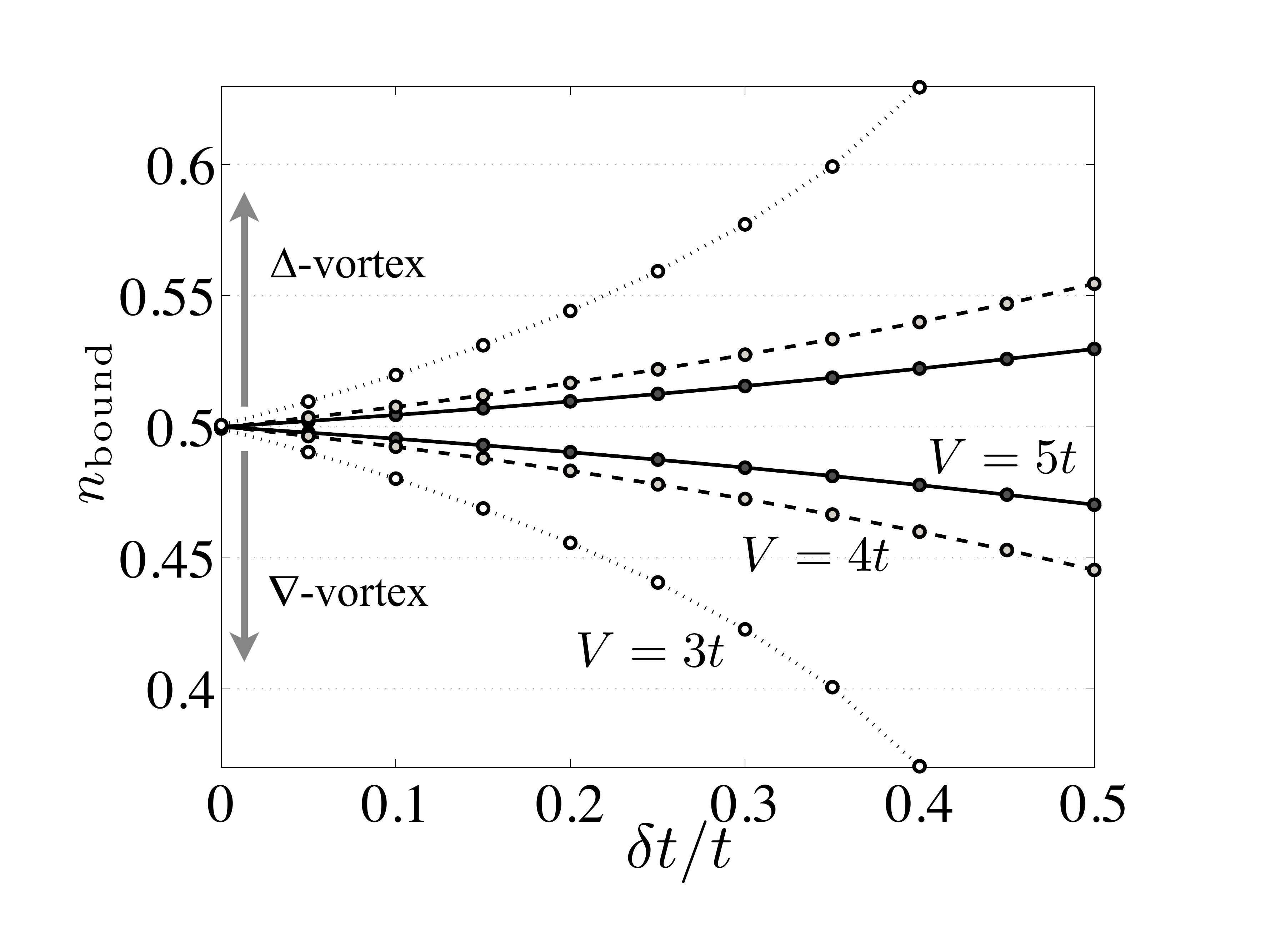}
\caption{The charge bound to a topological point defect as function of the symmetry-breaking field $\delta t$ for interactions $V=3t$, $V=4t$ and $V=5t$. The upper set of curves corresponds to the $\triangle$-vortices while the lower one corresponds to the $\triangledown$-vortices.}
\label{fig:irrat}
\end{figure}

%%%%%%%%%%%%%%%%%%%%%%%%%%
The value of $1/2$ for the charge bound to a topological defect depends on a crucial symmetry, namely that the up-triangles are equivalent to the down triangles. To see this we now consider the effect of a finite $\delta t$ in Eq.~\eqref{eq:HBOW}. This perturbation effectively increases the hopping on the up triangles, $t_{\triangle}=t+\delta t$, thereby breaking the symmetry between the up and the down triangles. We have calculated the charge bound to a defect as function of $\delta t$ for different interaction strength. The result is shown in Fig.~\ref{fig:irrat}. Clearly, the value of the charge varies continuously with the strength of the symmetry breaking potential $\delta t/t$. The effect is larger for smaller values of the interaction.
%%%%%%%
%%%%%%
This behavior is in agreement with Refs.~\onlinecite{Chamon:2008,Chamon:2008b} where field-theoretical methods have been used to study the effect of a symmetry-breaking potential.  Interestingly, it was found that by introducing a chiral gauge-field the energy of a single defect becomes finite turning them into well-defined excitations.\cite{Jackiw:2007} But if this happens, the fractional charge is rerationalized to 1/2.\cite{Chamon:2008} Turning this argument around, we may view the dependence of the charge on $\delta t$ as a manifestation of the fact that in our system a chiral gauge field is absent and that a single topological point defect costs an energy which depends on the system size.
We want to stress again that the value of $1/2$ is protected by the symmetry between up and down triangles and does {\it not} result from a spectral symmetry of the single-particle excitations. Such a particle-hole symmetry is absent on the kagome lattice and only emerges in the low-energy description, see Sec.~\ref{sec:Dirac}.
%%%%%%%%%%%%%
\subsection{Polaron state}
\label{sec:polaron}
%%%%%%%%%%%%%
%%%%%%%
\begin{figure}
\centering
\includegraphics[width=0.95\linewidth]{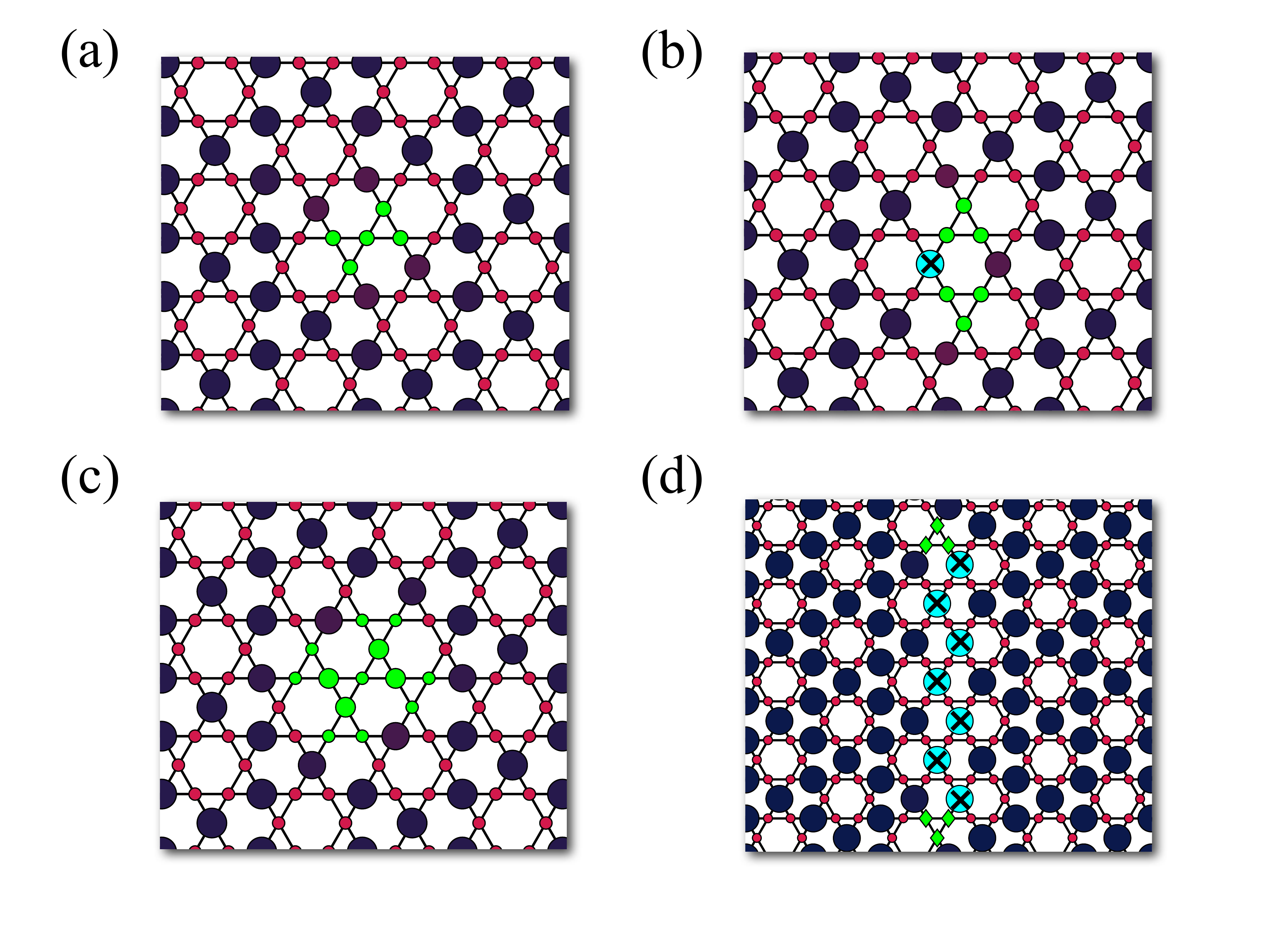}
\caption{(Color online.) The charge distribution of different polaron states obtained for $V=3t$. In (b) and (d), we have marked the misplaced Fermi-rich sites created by separating the up and down defects with a cross.}
\label{fig:holeconfig}
\end{figure} 
%%%%%%
In the previous section, the geometry and the boundary conditions have been chosen such that the property of an isolated defect can be studied. However, to have a configuration which only costs a finite energy in the thermodynamic limit, the vortex has to be healed at some point. This is possible when considering pairs of defects and naturally shows up when we study the property of a single hole doped into a large system. 

In the following, we want to find the ground state mean-field solution for a single hole. First, we note that the uniform mean-field solution which preserves the translational symmetry is always higher in energy than several solutions with inhomogeneous charge distributions where the inhomogeneity is restricted to a relatively small region. This signals the failure of the rigid band picture in the interacting system. As a matter of fact, a single hole doped into the uniform phase tends to polarize its surrounding which leads to an inhomogeneous charge distribution around the hole. Following standard nomenclature, we refer to the hole with its polarizing cloud as a {\it polaron state}. What is special in the present system is that the polaron has an internal structure. In fact, it can be viewed as a bound state of an up and a down defect. The confinement of the two defects results from the energy cost of domain walls which are necessarily created when trying to separate them in real space. 

Figure~\ref{fig:holeconfig} shows the self-consistent charge distribution of various polaron states for $V=3t$. The density distribution shown in (a) can be thought as the result of removing an electron from a Fermi-rich site. Clearly, the polaron wave function is well localized. Panel (c) again shows a fairly well-localized polaron state with large isotropy. However, in this case the polaron is in an excited state. Panel (b) shows the situation where the position of a Fermi-rich site (marked with a cross) has changed as compared to (a). As a result, up and down defects have been separated and the polaron wave function acquires two components. By moving the position of other Fermi-rich sites, the two defects can be separated even further, as shown in (d). In the simplest case, this procedure generates a straight ``string" which connects the two defects. If the charge-density wave is in the ground state A at infinity, the string of misplaced Fermi-rich sites can be viewed as the phases B and C with a minimal extension in the direction perpendicular to the string. (We note that it is also possible to stabilize solutions where all the three phases are extended but we have found that these configurations have higher energy than the straight string for the same separation of the two defects.) 
%%%%%%%
\begin{figure}[b]
\centering
\includegraphics[width=0.7\linewidth]{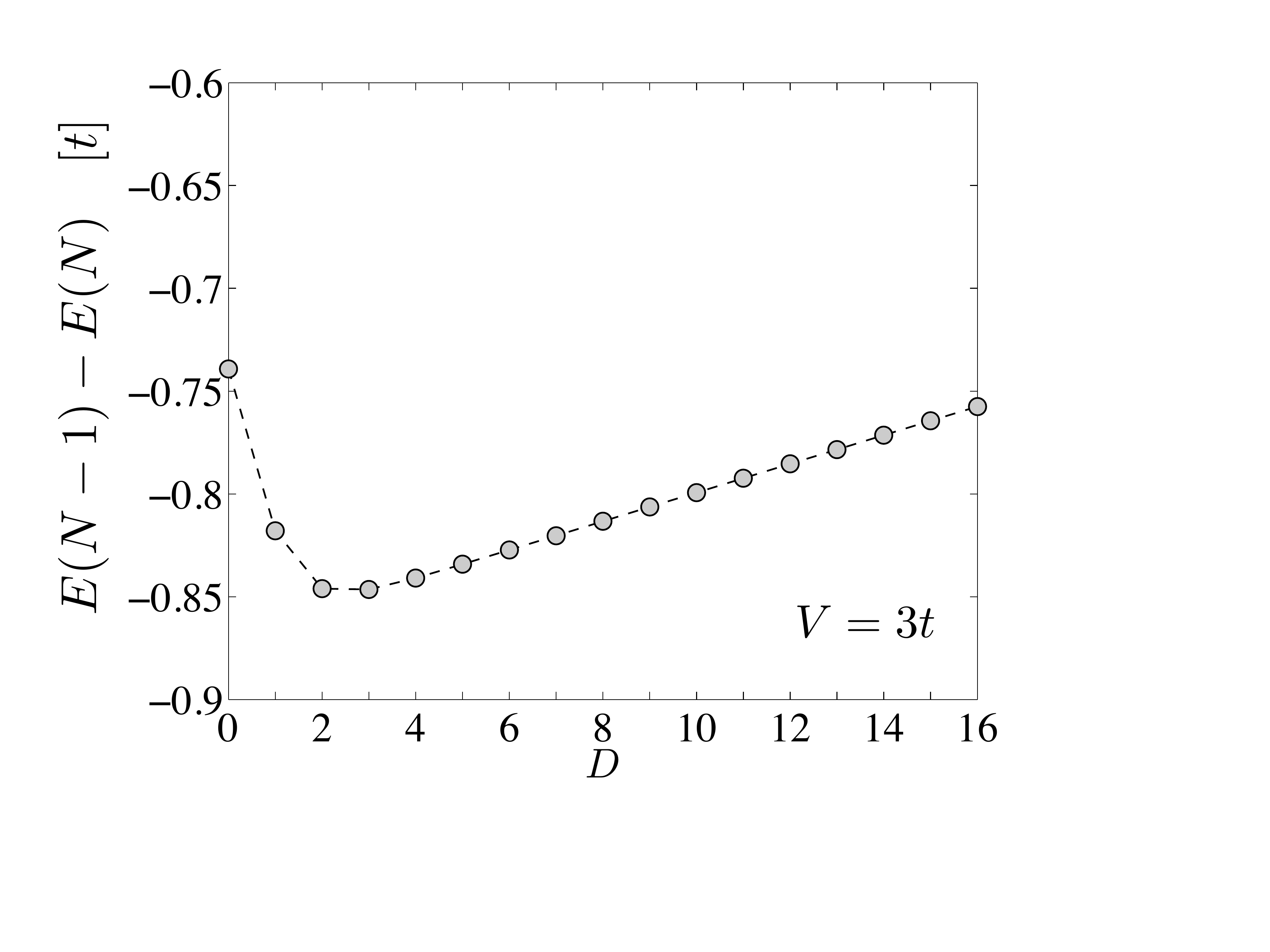}
\includegraphics[width=0.7\linewidth]{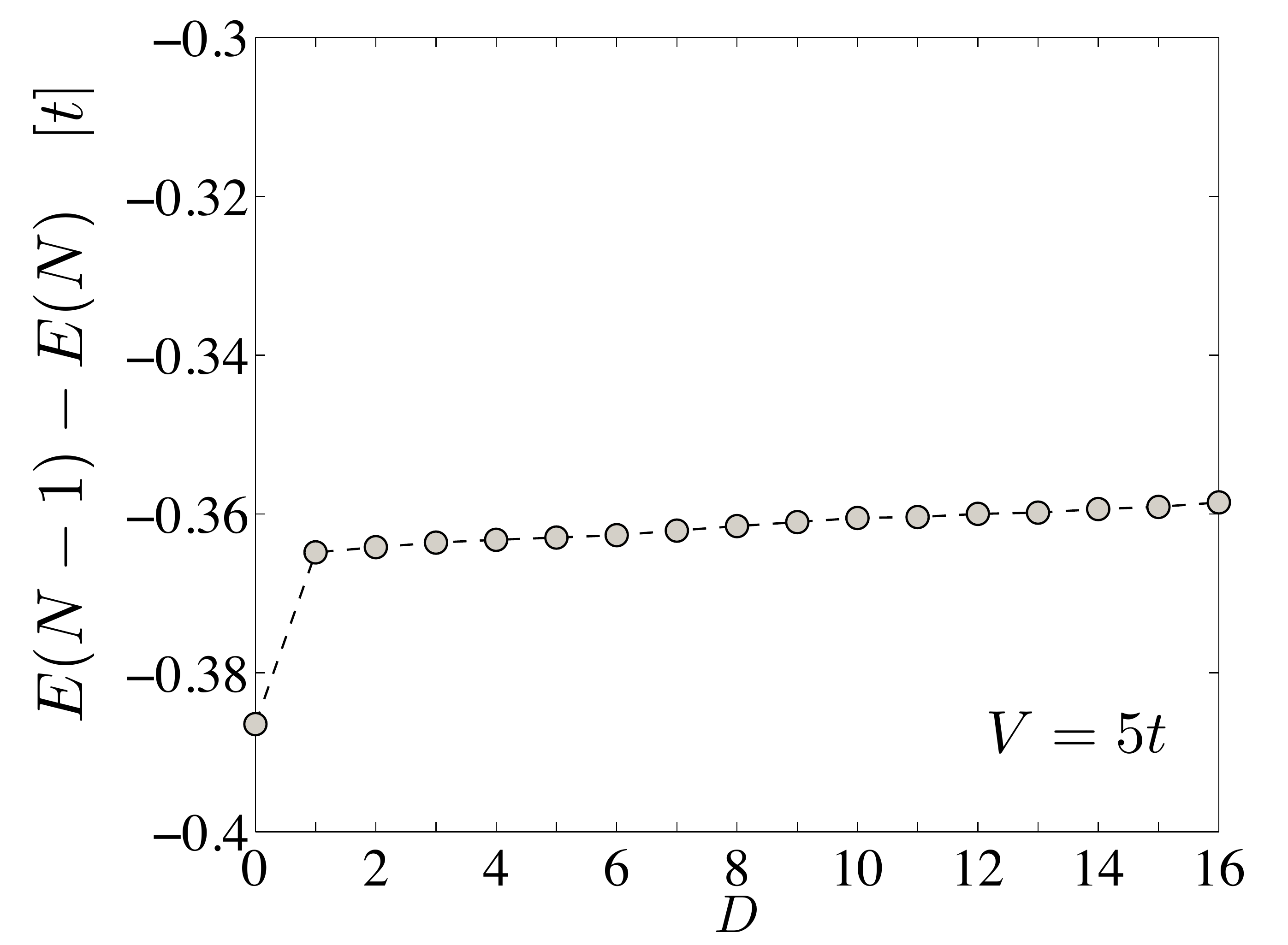}
\caption{The confinement potential between two fractional defects in the charge ordered kagome lattice. Shown is the energy difference between mean-field states with a single hole and the undoped uniform state as function of the number $D$ of misplaced sites necessary to separate the two defects, see Fig.~\ref{fig:holeconfig}. In the top panel, the nearest neighbor interaction is $V=3t$. The minimal energy occurs at a separation $D=3$. In the bottom panel, the nearest neighbor interaction of $V=5t$ is considered. The minimal energy occurs at zero separation of the defects. Results are obtained by solving the self-consistency equations on a hexagonal cluster with periodic boundary conditions including 1296 sites.}
\label{fig:confpot}
\end{figure}
%%%%%%
The energy as a function of the number $D$ of misplaced sites [the sites marked with a cross in Fig.~\ref{fig:holeconfig}(d)] measures the confinement between the two defects. Figure~\ref{fig:confpot} shows the energy of the hole as function of $D$ obtained for $V=3t$ and $V=5t$. In both cases, the energy grows linearly with $D$ for large $D$. This is in agreement with the expectation that every misplaced site costs the same energy because the ring exchange in the hexagons participating in the string is no longer effective.\cite{OBrien:2010} The slope is smaller for larger $V$ and eventually vanishes for $V/t\rightarrow\infty$. In this limit, the two defects are free to separate. For $V=5t$, the lowest energy configuration is the one with tightly bound defects [cf.~Fig.~\ref{fig:holeconfig}(a)]. It then costs a finite energy to separate the two defects by one unit [cf.~Fig.~\ref{fig:holeconfig}(b)] and after that the energy increases linearly with distance $D$. 
Interestingly, the situation looks quite different for weaker interactions. Namely, as shown in Fig.~\ref{fig:confpot} for $V=3t$, the lowest energy configuration corresponds to two defects separated by a string of length $D=3$. Thus, in this regime, the polaron state has a diatomic molecule character where the confinement length exceeds the defect size. 

Quantum mechanical processes which go beyond the static mean-field description (such as the ring exchange \cite{OBrien:2010}) will alter the quantitative dependence of the confining potential on $t/V$. However, we expect the qualitative aspects of the static mean-field solutions to be robust against a more careful treatment of these processes.

\begin{figure}
\centering
\includegraphics[width=0.8\linewidth]{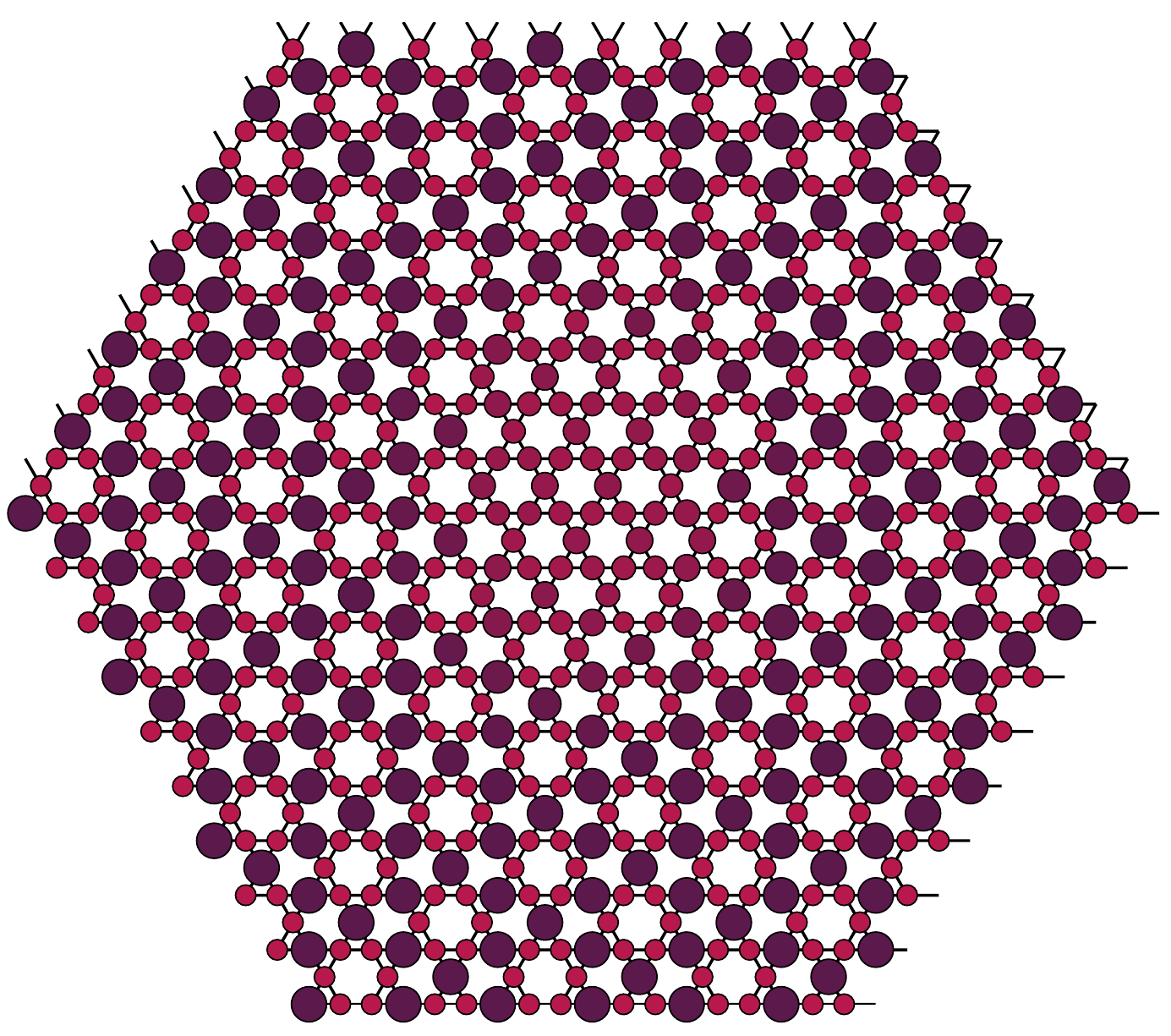}
\caption{(Color online.) The charge distribution of a polaron state in the weakly ordered charge-density wave. The interaction has been set to $V=2.3t$.}
\label{fig:polaronV2p3}
\end{figure}

The above systematic approach only works when the defect size is comparable to the lattice spacing which is the case for $V\gtrsim 3t$. We have also numerically studied the polaron wave function for smaller interactions in the weakly ordered state close to $V_c$. A typical converged solution is shown in Fig.~\ref{fig:polaronV2p3} for $V=2.3t$. In this regime, the polaron extends over several lattice spacings and it was not possible to control the location of the elementary up and down defects. Moreover, typical solutions are rather isotropic. This indicates that the confinement length at zero temperature is smaller than the defect size.
%%%%%%%
%%%%%%

In general, we expect that the polaron is dynamical and there is a center of mass motion as well as a relative motion of the two defects forming the polaron. For a clean system, it is likely that the polaron is not localized in real space. Rather, one would try to restore the translational symmetry by taking a superposition of localized polaron states with the same or nearby energies by considering the configuration interaction between the different mean-field states. Such an approach has for example been used to study the dispersion of a doped hole in the Hubbard model.\cite{Louis:1999} On the other hand, in the presence of imperfections, trapping of the polaron can occur. 

In the strongly correlated regime, the quantum mechanical polaron wave function has a large spatial extent because the confining potential is weak. Therefore, increasing the doping concentration could lead to new quantum phases where the confining is no longer relevant. For example, one can speculate that a plasma of fractionally charged defects is realized once the mean distance between polarons fall below the average diameter of a single bound pair.\cite{Runge:2007} Other interesting phases may involve crystalline structures of fractionally charged defects.

%%%%%%%%%%%%%%%%%%%%%%%%%%%%
\section{Weakly ordered state}
\label{sec:Dirac}
%%%%%%%%%%%%%%%%%%%%%%%%%%%%
To overcome the limitations of the numerical approach for the weakly ordered state we now turn to a more analytical description of topological point defects in the regime where both the order parameter and its gradient are small. Thereby, we are making a connection to the solitonic fractionalization mechanism in two dimensions.\cite{Mudry:2007} Thus, we assume that we are sufficiently close to the phase boundary ($V\gtrsim V_c$) such that an expansion in the order parameter is justified. In linear order, only the low energy degrees of freedom in the vicinity of the two Dirac points enter. 

%%%%%%%
\begin{figure}
\centering
\includegraphics[width=0.6\linewidth]{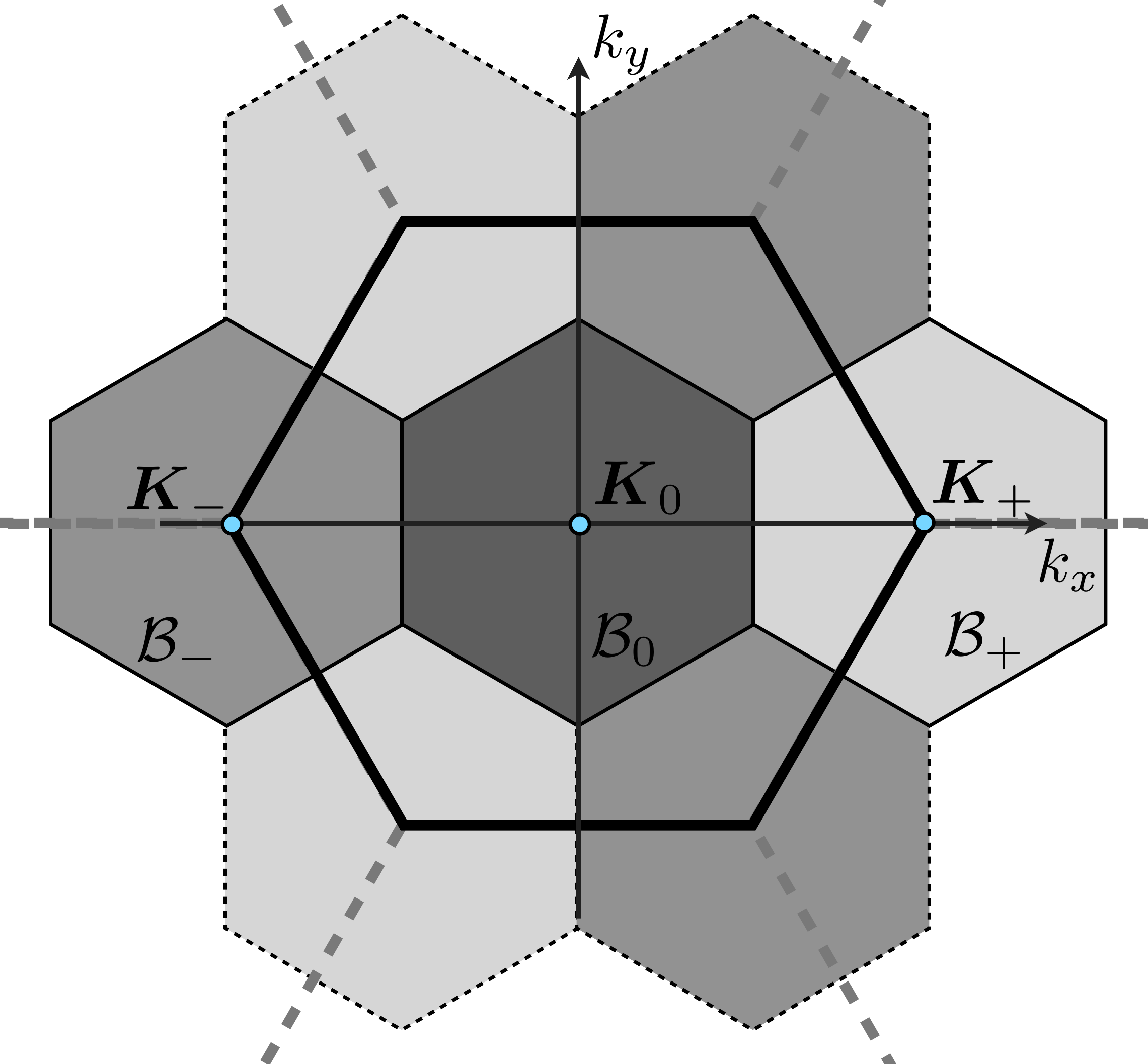}
\caption{The $\sqrt{3}\times\sqrt{3}$ reconstruction of the charge ordered state leads to a partitioning of the Brillouin zone.}
\label{fig:BZ}
\end{figure}
%%%%%%
For the analytical treatment it is convenient to use a notation which is adapted to the $\sqrt{3}\times\sqrt{3}$ reconstruction of the unit cell in the ordered state. Hence, we divide the first Brillouin zone into three patches ${\mathcal B}_0$ and ${\mathcal B}_\pm$ located around ${\bs \Gamma\equiv{\bs K}_0}=(0,0)$ and ${\bs K_{\pm}}$ as shown in Fig.~\ref{fig:BZ}. We will always use the convention that a capital $\bs K$ or $\bs Q$ is defined in the original Brillouin zone ${\mathcal B}=\oplus_l{\mathcal B}_l$ while a small $\bs k$ or $\bs q$ denotes a vector in ${\mathcal B}_0$. Then, for ${\bs K}\in{\mathcal B}_l$ $(l=0,\pm)$ we decompose the crystal momentum according to ${\bs K}={\bs K}_l+{\bs k}$ with ${\bs k}\in \mathcal{B}_0$ and define $c_{\nu l}({\bs k})=c_{\nu}({\bs K})$.
This allows us to write the local Fermi operators as a sum over the patches in the following way
\begin{equation}
c_{\nu}({\bs r})=\frac{1}{\sqrt{3}}\sum_l c_{\nu l}({\bs r})e^{i{\bs K}_l\cdot({\bs r}+{\bs r}_{\nu})}.
\end{equation}
Here, we have separated out the oscillatory factors $\exp[{\bs K}_l\cdot({\bs r}+{\bs r}_{\nu})]$ and have defined
\begin{equation}
c_{\nu l}({\bs r})=\sqrt{\frac{3}{N}}\sum_{{\bs k}\in{\mathcal B}_0}c_{\nu l}({\bs k})e^{i{\bs k}\cdot({\bs r}+{\bs r}_{\nu})}.
\end{equation}
The states in one patch live on a kagome lattice in real space with a threefold enlarged unit cell.
The single-particle operators in the different patches are related by the uniform ordering vector ${\bs G}$:
\begin{subequations}
\label{eq:shift}
\begin{align}
c_{\nu -}({\bs k}+{\bs G})&=c_{\nu +}({\bs k}),\\
c_{\nu 0}({\bs k}+{\bs G})&=c_{\nu -}({\bs k}),\\
c_{\nu +}({\bs k}+{\bs G})&=c_{\nu 0}({\bs k}).
\end{align}
\end{subequations}
Therefore, the Fourier components $\rho_{\alpha}({\bs Q})$ with ${\bs Q}$ close to $\pm {\bs G}$ couple the single particle states between different patches and most importantly, between the two Dirac cones.
\subsection{Effective sublattice basis on the kagome lattice}
For the low-energy description it is justified to truncate the Hilbert space by restricting to the single-particle states in the vicinity of the Dirac points at $\bs K_{\pm}$, see Eq.~\eqref{eq:Kpm}. Therefore, only operators associated with the two valleys $l=\pm$ are kept.
The next step involves a ${\bs k}$-independent transformation from the site to the ``sublattice" basis:
\begin{subequations}
\label{eq:trafo}
\begin{align}
c_{1,\pm}({\bs k})&=\frac{e^{\pm i\phi}}{\sqrt{3}}\left[e^{\mp i\pi/3}a_{\pm}({\bs k})+e^{\pm i\pi/3}b_{\pm}({\bs k})\right],\\
c_{2,\pm}({\bs k})&=\frac{e^{\pm i\phi}}{\sqrt{3}}\left[e^{\pm i\pi/3}a_{\pm}({\bs k})+e^{\mp i\pi/3}b_{\pm}({\bs k})\right],\\
c_{3,\pm}({\bs k})&=\frac{e^{\pm i\phi}}{\sqrt{3}}\left[a_{\pm}({\bs k})+b_{\pm}({\bs k})\right].
\end{align}
\end{subequations}
Above, we have suppressed the contribution of operators acting on states of the flat band at higher energy.  The $a$ and the $b$ operators are chosen in analogy to graphene in which case they would act on states living either on the A or the B sublattice. It turns out that on the kagome lattice, the up and the down triangles play the role of the A and B sites. This becomes clear when inverting the relation \eqref{eq:trafo} for $a_{\pm}^{\dag}(0)$ and $b_{\pm}^{\dag}(0)$ and expanding in terms of real space operators:\cite{Indergand:2006b}
\begin{subequations}
\begin{align}
a_{+}^{\dag}(0)&=\frac{1}{\sqrt{3N}}\sum_{\bs r}\left[\omega^2c_1^{\dag}({\bs r})\!+\!\omega c_2^{\dag}({\bs r})\!+\!c_3^{\dag}({\bs r})\right]e^{i{\bs K}_+\cdot{\bs r}},\\
b_{+}^{\dag}(0)&=\frac{1}{\sqrt{3N}}\sum_{\bs r}\left[c_1^{\dag}({\bs r})+c_2^{\dag}({\bs r})+c_3^{\dag}({\bs r})\right]e^{i{\bs K}_+\cdot{\bs r}},\\
a_{-}^{\dag}(0)&=\frac{1}{\sqrt{3N}}\sum_{\bs r}\left[\omega c_1^{\dag}({\bs r})\!+\!\omega^2c_2^{\dag}({\bs r})\!+\!c_3^{\dag}({\bs r})\right]e^{i{\bs K}_-\cdot{\bs r}},\\
b_{-}^{\dag}(0)&=\frac{1}{\sqrt{3N}}\sum_{\bs r}\left[c_1^{\dag}({\bs r})+c_2^{\dag}({\bs r})\!+\!c_3^{\dag}({\bs r})\right]e^{i{\bs K}_-\cdot{\bs r}}.
\end{align}
\label{eq:ab-real}
\end{subequations}
Here, we have introduced $\omega=\exp(2\pi i/3)$ and have set $\phi=0$ for clarity. In the state created by $a_{+}^{\dag}(0)$ $[a_{-}^{\dag}(0)]$, the phase on every up triangle increases by $2\pi/3$ along each bond in the [anti-]clockwise direction while the phase remains constant on the down triangles. On the other hand, in the state created by $b_{-}^{\dag}(0)$ $[b_{+}^{\dag}(0)]$, the phase on every down triangle increases by $2\pi/3$ along each bond in the [anti-]clockwise direction while the phase remains constant on the up triangles.
\subsection{Projected mean-field Hamiltonian}
The operators introduced in the previous section allows one to obtain an effective low-energy Hamiltonian. The calculation is straight forward but lengthy. In the following we will only present the final results.
\subsubsection{Kinetic energy}
Linearizing in ${\bs k}$ around ${\bs K}_{\pm}$ and applying the transformation \eqref{eq:trafo} brings the low energy tight-binding Hamiltonian into the canonical Dirac form
\begin{equation}
H_0=\sum_{l=\pm}\sum_{{\bs k}\in{\mathcal B}_0}\left[v_Fl(k_x+ilk_y)a_l^{\dag}({\bs k})b_l({\bs k})+{\rm h.c.}\right]
\label{eq:H0}
\end{equation}
with the Fermi velocity $v_F=\sqrt{3}ta/2$ ($\hbar\equiv1$).  We have shifted the zero of energy to the Dirac points.
\subsubsection{Bond order term}
The term Eq.~\eqref{eq:HBOW} which breaks the symmetry between the up and down triangles enters the low energy description as a staggered potential $\mu_s=-3/2\delta t$ for the effective sublattice states:
\begin{equation}
H_{\rm BOW}=-\mu_s\sum_{l=\pm}\sum_{{\bs k}\in{\mathcal B}_0}\left[a_l^{\dag}({\bs k})a_l({\bs k})-b_l^{\dag}({\bs k})b_l({\bs k})\right].
\end{equation}
Here, terms of order $\mathcal{O}(k^2)$ have been neglected. The above form is in full analogy with a staggered potential on the honeycomb lattice. Furthermore, the effect of a finite $\mu_s$ is plausible when considering the representation Eq.~\eqref{eq:ab-real}.
\subsubsection{Mean-field interaction}
In order to project this operator onto the low-energy degrees we first write the density operators $n_{\nu}(-{\bs Q})$ in terms of the patch operators $c_{\nu l}({\bs k})$. Then, using the transformation \eqref{eq:trafo} and keeping only operators which act on the low-energy degrees alone, the mean-field interaction assumes the following form
\begin{equation}
H_V'=\sum_{{\bs k},{\bs q}}\Phi^{\dag}({\bs k}-{\bs q})\hat{V}({\bs q})\Phi({\bs k})+{\rm const}.
\label{eq:HVred}
\end{equation}
Here, the summation is over the reduced Brillouin zone, ${\bs k}$ and ${\bs q}\in{\mathcal B}_0$. Furthermore, we have introduced the four-component spinor
\begin{equation}
\Phi({\bs k})\equiv \begin{pmatrix}
b_+({\bs k})\\
a_+({\bs k})\\
a_-({\bs k})\\
b_-({\bs k})
\end{pmatrix}=\sqrt{\frac{3}{N}}\sum_{{\bs R}}
\Phi({\bs R})e^{-i{\bs k}\cdot{\bs R}}.
\end{equation}
Note that $\Phi({\bs R})$ is a coarse grained operator defined on a triangular lattice with a three-fold bigger unit cell. We assume that both $|\Delta|$ and the gradient ${\vec \nabla}\varphi$ are small and we keep only terms entering linear in these quantities. For the matrix $\hat{V}({\bs q})$ we then find the following simple expression
\begin{equation}
\hat{V}({\bs q})=
\begin{pmatrix}
0 & \Delta({\bs q}){\bs 1}\\
\Delta^*({\bs q}) {\bs 1}& 0
\end{pmatrix}.
\end{equation}
Here, ${\bs 1}$ is the 2$\times$2 identity matrix and
\begin{equation}
\Delta({\bs q})=-\frac{2V}{3N}\left[\rho_1({\bs G}+{\bs q})+\rho_2({\bs G}+{\bs q})+\rho_3({\bs G}+{\bs q})\right]
\end{equation}
is assumed to be peaked around ${\bs q}\approx{\vec \nabla}\varphi$.
\subsubsection{Continuum limit}
The continuum limit is defined by $a\rightarrow0$ while keeping the Fermi velocity $v_F$ and the coupling constant $V'=Vv_{\rm uc}$ constant. It then follows that we scale the fermion fields according to
\begin{equation}
v_{\rm uc}\sum_{\bs R}\dots\rightarrow\int d^2{\bs R}\dots,\quad\Phi\rightarrow\Psi\equiv\Phi/\sqrt{v_{\rm uc}}.
\label{eq:Psi}
\end{equation}
where the unit cell volume is $v_{\rm uc}=3\sqrt{3}a^2/2$. The linearized mean-field theory takes the following continuum form
\begin{equation}
{\mathcal H}=\int \!\!d^2R\, \Big[\Psi^{\dag}({\bs R})\, \mathcal{K}\Psi({\bs R})+\frac{9}{2V'}|\Delta({\bs R})|^2\Big]
\label{eq:CG}
\end{equation}
for the four-component wave function $\Psi({\bs R})$ given in Eq.~\eqref{eq:Psi}. In the notation similar to Ref.~\onlinecite{Jackiw:2007} the kernel ${\mathcal K}$ in Eq.~\eqref{eq:CG} is written as
\begin{equation}
{\mathcal K}=v_F{\bs \alpha}\cdot\left(-i{\bs\nabla }\right)+\mu_s R+\beta\left[\Delta_1({\bs R})-i\gamma_5\Delta_2({\bs R})\right].
\label{eq:K}
\end{equation}
We used the $4\times4$ Dirac matrices
\begin{equation}
\alpha_i=\begin{pmatrix}
\sigma_i&0\\
0&-\sigma_i\end{pmatrix},\quad
\beta=\begin{pmatrix}
0&{\bs 1}\\
{\bs 1}&0
\end{pmatrix},\quad
\gamma_5=\begin{pmatrix}
{\bs 1}&0\\
0&-{\bs 1}
\end{pmatrix},
\end{equation}
where $i=x,y$ and the $2\times2$ matrices $\sigma_i$ and $R$ are defined as
\begin{equation}
\sigma_x=\begin{pmatrix}
0&1\\
1&0
\end{pmatrix},
\quad
\sigma_y=\begin{pmatrix}
0&-i\\
i&0
\end{pmatrix}\quad{\rm and}\quad
R\equiv\sigma_z=\begin{pmatrix}
1&0\\
0&-1
\end{pmatrix}.
\label{eq:Pauli}
\end{equation}
$\Delta_1$ and $\Delta_2$ are real and imaginary part of $\Delta$.
%%%
\subsection{Vortex solution and mid gap states}
We are now in a position to study the effect of a vortex in $\Delta({\bs R})$ on the fermionic spectrum. Thereby, we apply the results previously found for graphene.\cite{Mudry:2007,Chamon:2008,Chamon:2008b,Seradjeh:2008} We assume a symmetric vortex configuration with vorticity $q$ which in polar coordinates is written as
\begin{equation}
\Delta(r,\theta)\equiv\Delta_1+i\Delta_2=\Delta_0(r)e^{i q\theta+\alpha}.
\end{equation}
The amplitude $\Delta_0(r)$ vanishes for $r\rightarrow 0$ and assumes a constant value $\Delta_0(\infty)$ far away from the origin. It has been shown that such a configuration leads to a single mid-gap state at an energy $E=\mu_s$ or $E=-\mu_s$, depending on the sign of the vorticity $q$.\cite{Chamon:2008b} These solutions merge into a single zero-energy mode in the limit $\mu_s\rightarrow 0$. From the emergent spectral symmetry of the Dirac equation and the completeness of states in the absence and presence of a vortex it follows that both the valence and the conduction band transfer half a state to the zero-energy mode. As a result, a charge $\pm 1/2$ is bound to the vortex depending on if the zero-energy mode is occupied or not. These simple considerations break down for a finite $\mu_s$ because the emergent particle-hole symmetry of the single particle spectrum is violated. The calculation of the bound charge in this situation is more involved. We follow here the argumentation of Ref.~\onlinecite{Chamon:2008,Chamon:2008b} which makes use of the fact that a inhomogeneous static configuration of the three real fields $\mu_s$, $\Delta_1$ and $\Delta_2$ induce a fermionic charge density. This charge density is obtained from a perturbative treatment around the uniform solution:\cite{Goldstone:1981}
\begin{equation}
\rho(x,y)=\frac{1}{4\pi}{\vec n}\cdot(\partial_x{\vec n}\times\partial_y{\vec n}).
\label{eq:rho-n}
\end{equation}
Here, $\vec{n}$ is a unit vector defined as
\begin{equation}
\vec{n}=\frac{1}{\sqrt{\Delta_0^2+\mu_s^2}}\begin{pmatrix}
\Delta_1\\ \Delta_2\\\mu_s
\end{pmatrix}.
\end{equation}
The total charge bound to the vortex can be obtained by integrating the density over space
\begin{equation}
Q=\int\!dxdy\, \rho(x,y).
\label{eq:Q}
\end{equation}
From Eq.~\eqref{eq:rho-n} it follows that the integral Eq.~\eqref{eq:Q} measures the area (in units of $4\pi$) which is covered on the sphere by the unit vector $\vec{n}$ in the mapping $(x,y)\mapsto \vec{n}(x,y)$. The result is therefore
\begin{equation}
Q=\frac{q}{2}\left[{\rm  sign}(\mu_s)-\frac{\mu_s}{\sqrt{\Delta_0(\infty)^2+\mu_s^2}}\right].
\end{equation}
In the limit $\mu_s\rightarrow 0^{\pm}$ and for $q=\pm1$ we recover the value $Q=\pm1/2$. On the other hand, the dependence of the charge on $\mu_s$ is similar to the result reported in Fig.~\ref{fig:irrat}.
%%%%%%%%%%%%
\section{Conclusions}
%%%%%%%%%%%%
We have studied topological point defects in the charge-density wave realized in a model of interacting spin-polarized fermions on the kagome lattice at filling fraction 1/3. We have found that elementary point defects carry a charge $\pm 1/2$ as long as the symmetry between the up and the down triangles of the kagome lattice is preserved. If this symmetry is violated, the bound charge varies continuously with the symmetry breaking term. Moreover, we have argued that in the classical limit the point defect corresponds to a local violation of the triangle rule and is therefore related to the fact that the interaction is frustrated. On the other hand, in the weakly ordered state, we made a connection to the solitonic fractionalization mechanism based on the Dirac equation with a vortex in the background field. The considered system therefore offers a unique possibility to realize these two different routes to fractionalization in the same model.

Using unrestricted mean-field calculations we have studied the ground state of a single hole doped into the charge-density wave. We have found that the polaron state can be viewed as a bound state of two defects both carrying a charge $1/2$. We have also calculated the confining potential between these two defects and have found that in the intermediate interaction regime it is minimized for a finite separation.

We note here that the charge ordered state considered in this article bears some similarity with the ``trimerized" phase considered previously in that it shares the same enlarged unit cell and also has three different ground states.\cite{Guo:2009,Wang:2010} However, the charge density-wave order seems to be more easily realized in an interacting system. Our basic conclusions remain valid when the spin degree of freedom is taken into account as long as the interaction still favors the charge-density wave. Nevertheless, in the spinfull case it is not the charge which fractionalizes but the defects carry either a spin 1/2 and no charge or a charge $\pm1$ and no spin.

\begin{acknowledgments}
We acknowledge stimulating discussions and correspondence with M.~Franz, S.~D.~Huber, M.~Kargarian, E.~Louis, and J.~Wen. We acknowledge financial support from ARO under Grant No. W911NF-09-1-0527 and NSF under Grant No. DMR-0955778.
\end{acknowledgments}

\bibliography{biblio}

%Merlin.mbs v4.21 2009-07-09.
\begin{thebibliography}{10}%
\makeatletter
\providecommand \@ifxundefined [1]{%
 \ifx #1\undefined \expandafter \@firstoftwo
 \else \expandafter \@secondoftwo
\fi
}%
\providecommand \@ifnum [1]{%
 \ifnum #1\expandafter \@firstoftwo
 \else \expandafter \@secondoftwo
\fi
}%
\providecommand \enquote [1]{``#1''}%
\providecommand \bibnamefont  [1]{#1}%
\providecommand \bibfnamefont [1]{#1}%
\providecommand \citenamefont [1]{#1}%
\providecommand\href[0]{\@sanitize\@href}%
\providecommand\@href[1]{\endgroup\@@startlink{#1}\endgroup\@@href}%
\providecommand\@@href[1]{#1\@@endlink}%
\providecommand \@sanitize [0]{\begingroup\catcode`\&12\catcode`\#12\relax}%
\@ifxundefined \pdfoutput {\@firstoftwo}{%
 \@ifnum{\z@=\pdfoutput}{\@firstoftwo}{\@secondoftwo}%
}{%
 \providecommand\@@startlink[1]{\leavevmode\special{html:<a href="#1">}}%
 \providecommand\@@endlink[0]{\special{html:</a>}}%
}{%
 \providecommand\@@startlink[1]{%
  \leavevmode
  \pdfstartlink
   attr{/Border[0 0 1 ]/H/I/C[0 1 1]}%
   user{/Subtype/Link/A<</Type/Action/S/URI/URI(#1)>>}%
  \relax
 }%
 \providecommand\@@endlink[0]{\pdfendlink}%
}%
\providecommand \url  [0]{\begingroup\@sanitize \@url }%
\providecommand \@url [1]{\endgroup\@href {#1}{\urlprefix}}%
\providecommand \urlprefix [0]{URL }%
\providecommand \Eprint[0]{\href }%
\@ifxundefined \urlstyle {%
  \providecommand \doi [1]{doi:\discretionary{}{}{}#1}%
}{%
  \providecommand \doi [0]{doi:\discretionary{}{}{}\begingroup
  \urlstyle{rm}\Url }%
}%
\providecommand \doibase [0]{http://dx.doi.org/}%
\providecommand \Doi[1]{\href{\doibase#1}}%
\providecommand \bibAnnote [3]{%
  \BibitemShut{#1}%
  \begin{quotation}\noindent
    \textsc{Key:}\ #2\\\textsc{Annotation:}\ #3%
  \end{quotation}%
}%
\providecommand \bibAnnoteFile [2]{%
  \IfFileExists{#2}{\bibAnnote {#1} {#2} {\input{#2}}}{}%
}%
\providecommand \typeout [0]{\immediate \write \m@ne }%
\providecommand \selectlanguage [0]{\@gobble}%
\providecommand \bibinfo [0]{\@secondoftwo}%
\providecommand \bibfield [0]{\@secondoftwo}%
\providecommand \translation [1]{[#1]}%
\providecommand \BibitemOpen[0]{}%
\providecommand \bibitemStop [0]{}%
\providecommand \bibitemNoStop [0]{.\EOS\space}%
\providecommand \EOS [0]{\spacefactor3000\relax}%
\providecommand \BibitemShut [1]{\csname bibitem#1\endcsname}%
%</preamble>
\bibitem{Su:1979}%
  \BibitemOpen
  \bibfield{author}{%
  \bibinfo {author} {\bibfnamefont{W.~P.}\ \bibnamefont{Su}}, \bibinfo {author}
  {\bibfnamefont{J.~R.}\ \bibnamefont{Schrieffer}},\ and\ \bibinfo {author}
  {\bibfnamefont{A.~J.}\ \bibnamefont{Heeger}},\ }%
  \bibfield{journal}{%
  \Doi{10.1103/PhysRevLett.42.1698}{\bibinfo {journal} {Phys. Rev. Lett.}}\ }%
  \textbf{\bibinfo {volume} {42}},\ \bibinfo {pages} {1698} (\bibinfo {year}
  {1979})%
  \bibAnnoteFile{NoStop}{Su:1979}%
\bibitem{Laughlin:1999}%
  \BibitemOpen
  \bibfield{author}{%
  \bibinfo {author} {\bibfnamefont{R.~B.}\ \bibnamefont{Laughlin}},\ }%
  \bibfield{journal}{%
  \Doi{10.1103/RevModPhys.71.863}{\bibinfo {journal} {Rev. Mod. Phys.}}\ }%
  \textbf{\bibinfo {volume} {71}},\ \bibinfo {pages} {863} (\bibinfo {year}
  {1999})%
  \bibAnnoteFile{NoStop}{Laughlin:1999}%
\bibitem{Castelnovo:2008}%
  \BibitemOpen
  \bibfield{author}{%
  \bibinfo {author} {\bibfnamefont{C.}~\bibnamefont{Castelnovo}}, \bibinfo
  {author} {\bibfnamefont{R.}~\bibnamefont{Moessner}},\ and\ \bibinfo {author}
  {\bibfnamefont{S.~L.}\ \bibnamefont{Sondhi}},\ }%
  \bibfield{journal}{%
  \Doi{10.1038/nature06433}{\bibinfo {journal} {Nature}}\ }%
  \textbf{\bibinfo {volume} {451}},\ \bibinfo {pages} {42} (\bibinfo {year}
  {2008})%
  \bibAnnoteFile{NoStop}{Castelnovo:2008}%
\bibitem{Oshikawa:2006}%
  \BibitemOpen
  \bibfield{author}{%
  \bibinfo {author} {\bibfnamefont{M.}~\bibnamefont{Oshikawa}}\ and\ \bibinfo
  {author} {\bibfnamefont{T.}~\bibnamefont{Senthil}},\ }%
  \bibfield{journal}{%
  \Doi{10.1103/PhysRevLett.96.060601}{\bibinfo {journal} {Phys. Rev. Lett.}}\
  }%
  \textbf{\bibinfo {volume} {96}},\ \bibinfo {pages} {060601} (\bibinfo {year}
  {2006})%
  \bibAnnoteFile{NoStop}{Oshikawa:2006}%
\bibitem{Balents:2002}%
  \BibitemOpen
  \bibfield{author}{%
  \bibinfo {author} {\bibfnamefont{L.}~\bibnamefont{Balents}}, \bibinfo
  {author} {\bibfnamefont{M.~P.~A.}\ \bibnamefont{Fisher}},\ and\ \bibinfo
  {author} {\bibfnamefont{S.~M.}\ \bibnamefont{Girvin}},\ }%
  \bibfield{journal}{%
  \Doi{10.1103/PhysRevB.65.224412}{\bibinfo {journal} {Phys. Rev. B}}\ }%
  \textbf{\bibinfo {volume} {65}},\ \bibinfo {pages} {224412} (\bibinfo {year}
  {2002})%
  \bibAnnoteFile{NoStop}{Balents:2002}%
\bibitem{Hermele:2004}%
  \BibitemOpen
  \bibfield{author}{%
  \bibinfo {author} {\bibfnamefont{M.}~\bibnamefont{Hermele}}, \bibinfo
  {author} {\bibfnamefont{M.~P.~A.}\ \bibnamefont{Fisher}},\ and\ \bibinfo
  {author} {\bibfnamefont{L.}~\bibnamefont{Balents}},\ }%
  \bibfield{journal}{%
  \Doi{10.1103/PhysRevB.69.064404}{\bibinfo {journal} {Phys. Rev. B}}\ }%
  \textbf{\bibinfo {volume} {69}},\ \bibinfo {pages} {064404} (\bibinfo {year}
  {2004})%
  \bibAnnoteFile{NoStop}{Hermele:2004}%
\bibitem{Banerjee:2008}%
  \BibitemOpen
  \bibfield{author}{%
  \bibinfo {author} {\bibfnamefont{A.}~\bibnamefont{Banerjee}}, \bibinfo
  {author} {\bibfnamefont{S.~V.}\ \bibnamefont{Isakov}}, \bibinfo {author}
  {\bibfnamefont{K.}~\bibnamefont{Damle}},\ and\ \bibinfo {author}
  {\bibfnamefont{Y.~B.}\ \bibnamefont{Kim}},\ }%
  \bibfield{journal}{%
  \Doi{10.1103/PhysRevLett.100.047208}{\bibinfo {journal} {Phys. Rev. Lett.}}\
  }%
  \textbf{\bibinfo {volume} {100}},\ \bibinfo {pages} {047208} (\bibinfo {year}
  {2008})%
  \bibAnnoteFile{NoStop}{Banerjee:2008}%
\bibitem{Senthil:2000}%
  \BibitemOpen
  \bibfield{author}{%
  \bibinfo {author} {\bibfnamefont{T.}~\bibnamefont{Senthil}}\ and\ \bibinfo
  {author} {\bibfnamefont{M.~P.~A.}\ \bibnamefont{Fisher}},\ }%
  \bibfield{journal}{%
  \Doi{10.1103/PhysRevB.62.7850}{\bibinfo {journal} {Phys. Rev. B}}\ }%
  \textbf{\bibinfo {volume} {62}},\ \bibinfo {pages} {7850} (\bibinfo {year}
  {2000})%
  \bibAnnoteFile{NoStop}{Senthil:2000}%
\bibitem{Jackiw:1976}%
  \BibitemOpen
  \bibfield{author}{%
  \bibinfo {author} {\bibfnamefont{R.}~\bibnamefont{Jackiw}}\ and\ \bibinfo
  {author} {\bibfnamefont{C.}~\bibnamefont{Rebbi}},\ }%
  \bibfield{journal}{%
  \Doi{10.1103/PhysRevD.13.3398}{\bibinfo {journal} {Phys. Rev. D}}\ }%
  \textbf{\bibinfo {volume} {13}},\ \bibinfo {pages} {3398} (\bibinfo {year}
  {1976})%
  \bibAnnoteFile{NoStop}{Jackiw:1976}%
\bibitem{Goldstone:1981}%
  \BibitemOpen
  \bibfield{author}{%
  \bibinfo {author} {\bibfnamefont{J.}~\bibnamefont{Goldstone}}\ and\ \bibinfo
  {author} {\bibfnamefont{F.}~\bibnamefont{Wilczek}},\ }%
  \bibfield{journal}{%
  \Doi{10.1103/PhysRevLett.47.986}{\bibinfo {journal} {Phys. Rev. Lett.}}\ }%
  \textbf{\bibinfo {volume} {47}},\ \bibinfo {pages} {986} (\bibinfo {year}
  {1981})%
  \bibAnnoteFile{NoStop}{Goldstone:1981}%
\bibitem{Mudry:2007}%
  \BibitemOpen
  \bibfield{author}{%
  \bibinfo {author} {\bibfnamefont{C.-Y.}\ \bibnamefont{Hou}}, \bibinfo
  {author} {\bibfnamefont{C.}~\bibnamefont{Chamon}},\ and\ \bibinfo {author}
  {\bibfnamefont{C.}~\bibnamefont{Mudry}},\ }%
  \bibfield{journal}{%
  \Doi{10.1103/PhysRevLett.98.186809}{\bibinfo {journal} {Phys. Rev. Lett.}}\
  }%
  \textbf{\bibinfo {volume} {98}},\ \bibinfo {pages} {186809} (\bibinfo {year}
  {2007})%
  \bibAnnoteFile{NoStop}{Mudry:2007}%
\bibitem{Jackiw:2007}%
  \BibitemOpen
  \bibfield{author}{%
  \bibinfo {author} {\bibfnamefont{R.}~\bibnamefont{Jackiw}}\ and\ \bibinfo
  {author} {\bibfnamefont{S.-Y.}\ \bibnamefont{Pi}},\ }%
  \bibfield{journal}{%
  \Doi{10.1103/PhysRevLett.98.266402}{\bibinfo {journal} {Phys. Rev. Lett.}}\
  }%
  \textbf{\bibinfo {volume} {98}},\ \bibinfo {pages} {266402} (\bibinfo {year}
  {2007})%
  \bibAnnoteFile{NoStop}{Jackiw:2007}%
\bibitem{Chamon:2008}%
  \BibitemOpen
  \bibfield{author}{%
  \bibinfo {author} {\bibfnamefont{C.}~\bibnamefont{Chamon}}, \bibinfo {author}
  {\bibfnamefont{C.-Y.}\ \bibnamefont{Hou}}, \bibinfo {author}
  {\bibfnamefont{R.}~\bibnamefont{Jackiw}}, \bibinfo {author}
  {\bibfnamefont{C.}~\bibnamefont{Mudry}}, \bibinfo {author}
  {\bibfnamefont{S.-Y.}\ \bibnamefont{Pi}},\ and\ \bibinfo {author}
  {\bibfnamefont{A.~P.}\ \bibnamefont{Schnyder}},\ }%
  \bibfield{journal}{%
  \Doi{10.1103/PhysRevLett.100.110405}{\bibinfo {journal} {Phys. Rev. Lett.}}\
  }%
  \textbf{\bibinfo {volume} {100}},\ \bibinfo {pages} {110405} (\bibinfo {year}
  {2008})%
  \bibAnnoteFile{NoStop}{Chamon:2008}%
\bibitem{Chamon:2008b}%
  \BibitemOpen
  \bibfield{author}{%
  \bibinfo {author} {\bibfnamefont{C.}~\bibnamefont{Chamon}}, \bibinfo {author}
  {\bibfnamefont{C.-Y.}\ \bibnamefont{Hou}}, \bibinfo {author}
  {\bibfnamefont{R.}~\bibnamefont{Jackiw}}, \bibinfo {author}
  {\bibfnamefont{C.}~\bibnamefont{Mudry}}, \bibinfo {author}
  {\bibfnamefont{S.-Y.}\ \bibnamefont{Pi}},\ and\ \bibinfo {author}
  {\bibfnamefont{G.}~\bibnamefont{Semenoff}},\ }%
  \bibfield{journal}{%
  \Doi{10.1103/PhysRevB.77.235431}{\bibinfo {journal} {Phys. Rev. B}}\ }%
  \textbf{\bibinfo {volume} {77}},\ \bibinfo {pages} {235431} (\bibinfo {year}
  {2008})%
  \bibAnnoteFile{NoStop}{Chamon:2008b}%
\bibitem{Seradjeh:2008b}%
  \BibitemOpen
  \bibfield{author}{%
  \bibinfo {author} {\bibfnamefont{B.}~\bibnamefont{Seradjeh}}, \bibinfo
  {author} {\bibfnamefont{C.}~\bibnamefont{Weeks}},\ and\ \bibinfo {author}
  {\bibfnamefont{M.}~\bibnamefont{Franz}},\ }%
  \bibfield{journal}{%
  \Doi{10.1103/PhysRevB.77.033104}{\bibinfo {journal} {Phys. Rev. B}}\ }%
  \textbf{\bibinfo {volume} {77}},\ \bibinfo {pages} {033104} (\bibinfo {year}
  {2008})%
  \bibAnnoteFile{NoStop}{Seradjeh:2008b}%
\bibitem{Guo:2009}%
  \BibitemOpen
  \bibfield{author}{%
  \bibinfo {author} {\bibfnamefont{H.-M.}\ \bibnamefont{Guo}}\ and\ \bibinfo
  {author} {\bibfnamefont{M.}~\bibnamefont{Franz}},\ }%
  \bibfield{journal}{%
  \Doi{10.1103/PhysRevB.80.113102}{\bibinfo {journal} {Phys. Rev. B}}\ }%
  \textbf{\bibinfo {volume} {80}},\ \bibinfo {eid} {113102} (\bibinfo {year}
  {2009})%
  \bibAnnoteFile{NoStop}{Guo:2009}%
\bibitem{Hou:2010}%
  \BibitemOpen
  \bibfield{author}{%
  \bibinfo {author} {\bibfnamefont{C.-Y.}\ \bibnamefont{Hou}}, \bibinfo
  {author} {\bibfnamefont{C.}~\bibnamefont{Chamon}},\ and\ \bibinfo {author}
  {\bibfnamefont{C.}~\bibnamefont{Mudry}},\ }%
  \bibfield{journal}{%
  \Doi{10.1103/PhysRevB.81.075427}{\bibinfo {journal} {Phys. Rev. B}}\ }%
  \textbf{\bibinfo {volume} {81}},\ \bibinfo {pages} {075427} (\bibinfo {year}
  {2010})%
  \bibAnnoteFile{NoStop}{Hou:2010}%
\bibitem{Liu:2010}%
  \BibitemOpen
  \bibfield{author}{%
  \bibinfo {author} {\bibfnamefont{X.}~\bibnamefont{Liu}}\ and\ \bibinfo
  {author} {\bibfnamefont{R.}~\bibnamefont{Zhang}},\ }%
  \bibfield{journal}{%
  \Doi{10.1016/j.aop.2009.10.004}{\bibinfo {journal} {Annals of Physics}}\ }%
  \textbf{\bibinfo {volume} {325}},\ \bibinfo {pages} {384 } (\bibinfo {year}
  {2010})%
  \bibAnnoteFile{NoStop}{Liu:2010}%
\bibitem{Weeks:2010}%
  \BibitemOpen
  \bibfield{author}{%
  \bibinfo {author} {\bibfnamefont{C.}~\bibnamefont{Weeks}}\ and\ \bibinfo
  {author} {\bibfnamefont{M.}~\bibnamefont{Franz}},\ }%
  \bibfield{journal}{%
  \Doi{10.1103/PhysRevB.81.085105}{\bibinfo {journal} {Phys. Rev. B}}\ }%
  \textbf{\bibinfo {volume} {81}},\ \bibinfo {pages} {085105} (\bibinfo {year}
  {2010})%
  \bibAnnoteFile{NoStop}{Weeks:2010}%
\bibitem{Wang:2010}%
  \BibitemOpen
  \bibfield{author}{%
  \bibinfo {author} {\bibfnamefont{Z.}~\bibnamefont{Wang}}\ and\ \bibinfo
  {author} {\bibfnamefont{P.}~\bibnamefont{Zhang}},\ }%
  \bibfield{journal}{%
  \Doi{10.1088/1367-2630/12/4/043055}{\bibinfo {journal} {New Journal of
  Physics}}\ }%
  \textbf{\bibinfo {volume} {12}},\ \bibinfo {pages} {043055} (\bibinfo {year}
  {2010})%
  \bibAnnoteFile{NoStop}{Wang:2010}%
\bibitem{Fulde:2002}%
  \BibitemOpen
  \bibfield{author}{%
  \bibinfo {author} {\bibfnamefont{P.}~\bibnamefont{Fulde}}, \bibinfo {author}
  {\bibfnamefont{K.}~\bibnamefont{Penc}},\ and\ \bibinfo {author}
  {\bibfnamefont{N.}~\bibnamefont{Shannon}},\ }%
  \bibfield{journal}{%
  \Doi{10.1002/1521-3889(200212)11:12<892::AID-ANDP892>3.0.CO;2-J}{\bibinfo
  {journal} {Annalen der Physik}}\ }%
  \textbf{\bibinfo {volume} {11}},\ \bibinfo {pages} {892} (\bibinfo {year}
  {2002})%
  \bibAnnoteFile{NoStop}{Fulde:2002}%
\bibitem{Runge:2004}%
  \BibitemOpen
  \bibfield{author}{%
  \bibinfo {author} {\bibfnamefont{E.}~\bibnamefont{Runge}}\ and\ \bibinfo
  {author} {\bibfnamefont{P.}~\bibnamefont{Fulde}},\ }%
  \bibfield{journal}{%
  \Doi{10.1103/PhysRevB.70.245113}{\bibinfo {journal} {Phys. Rev. B}}\ }%
  \textbf{\bibinfo {volume} {70}},\ \bibinfo {pages} {245113} (\bibinfo {year}
  {2004})%
  \bibAnnoteFile{NoStop}{Runge:2004}%
\bibitem{Rokhsar:1988}%
  \BibitemOpen
  \bibfield{author}{%
  \bibinfo {author} {\bibfnamefont{D.~S.}\ \bibnamefont{Rokhsar}}\ and\
  \bibinfo {author} {\bibfnamefont{S.~A.}\ \bibnamefont{Kivelson}},\ }%
  \bibfield{journal}{%
  \Doi{10.1103/PhysRevLett.61.2376}{\bibinfo {journal} {Phys. Rev. Lett.}}\ }%
  \textbf{\bibinfo {volume} {61}},\ \bibinfo {pages} {2376} (\bibinfo {year}
  {1988})%
  \bibAnnoteFile{NoStop}{Rokhsar:1988}%
\bibitem{Moessner:2001a}%
  \BibitemOpen
  \bibfield{author}{%
  \bibinfo {author} {\bibfnamefont{R.}~\bibnamefont{Moessner}}\ and\ \bibinfo
  {author} {\bibfnamefont{S.~L.}\ \bibnamefont{Sondhi}},\ }%
  \bibfield{journal}{%
  \Doi{10.1103/PhysRevLett.86.1881}{\bibinfo {journal} {Phys. Rev. Lett.}}\ }%
  \textbf{\bibinfo {volume} {86}},\ \bibinfo {pages} {1881} (\bibinfo {year}
  {2001})%
  \bibAnnoteFile{NoStop}{Moessner:2001a}%
\bibitem{Fradkin:2004}%
  \BibitemOpen
  \bibfield{author}{%
  \bibinfo {author} {\bibfnamefont{E.}~\bibnamefont{Fradkin}}, \bibinfo
  {author} {\bibfnamefont{D.~A.}\ \bibnamefont{Huse}}, \bibinfo {author}
  {\bibfnamefont{R.}~\bibnamefont{Moessner}}, \bibinfo {author}
  {\bibfnamefont{V.}~\bibnamefont{Oganesyan}},\ and\ \bibinfo {author}
  {\bibfnamefont{S.~L.}\ \bibnamefont{Sondhi}},\ }%
  \bibfield{journal}{%
  \Doi{10.1103/PhysRevB.69.224415}{\bibinfo {journal} {Phys. Rev. B}}\ }%
  \textbf{\bibinfo {volume} {69}},\ \bibinfo {pages} {224415} (\bibinfo {year}
  {2004})%
  \bibAnnoteFile{NoStop}{Fradkin:2004}%
\bibitem{Sikora:2009}%
  \BibitemOpen
  \bibfield{author}{%
  \bibinfo {author} {\bibfnamefont{O.}~\bibnamefont{Sikora}}, \bibinfo {author}
  {\bibfnamefont{F.}~\bibnamefont{Pollmann}}, \bibinfo {author}
  {\bibfnamefont{N.}~\bibnamefont{Shannon}}, \bibinfo {author}
  {\bibfnamefont{K.}~\bibnamefont{Penc}},\ and\ \bibinfo {author}
  {\bibfnamefont{P.}~\bibnamefont{Fulde}},\ }%
  \bibfield{journal}{%
  \Doi{10.1103/PhysRevLett.103.247001}{\bibinfo {journal} {Phys. Rev. Lett.}}\
  }%
  \textbf{\bibinfo {volume} {103}},\ \bibinfo {pages} {247001} (\bibinfo {year}
  {2009})%
  \bibAnnoteFile{NoStop}{Sikora:2009}%
\bibitem{Wen:2010}%
  \BibitemOpen
  \bibfield{author}{%
  \bibinfo {author} {\bibfnamefont{J.}~\bibnamefont{Wen}}, \bibinfo {author}
  {\bibfnamefont{A.}~\bibnamefont{R\"uegg}}, \bibinfo {author}
  {\bibfnamefont{C.-C.~J.}\ \bibnamefont{Wang}},\ and\ \bibinfo {author}
  {\bibfnamefont{G.~A.}\ \bibnamefont{Fiete}},\ }%
  \bibfield{journal}{%
  \Doi{10.1103/PhysRevB.82.075125}{\bibinfo {journal} {Phys. Rev. B}}\ }%
  \textbf{\bibinfo {volume} {82}},\ \bibinfo {pages} {075125} (\bibinfo {year}
  {2010})%
  \bibAnnoteFile{NoStop}{Wen:2010}%
\bibitem{Nishimoto:2010}%
  \BibitemOpen
  \bibfield{author}{%
  \bibinfo {author} {\bibfnamefont{S.}~\bibnamefont{Nishimoto}}, \bibinfo
  {author} {\bibfnamefont{M.}~\bibnamefont{Nakamura}}, \bibinfo {author}
  {\bibfnamefont{A.}~\bibnamefont{O'Brien}},\ and\ \bibinfo {author}
  {\bibfnamefont{P.}~\bibnamefont{Fulde}},\ }%
  \bibfield{journal}{%
  \Doi{10.1103/PhysRevLett.104.196401}{\bibinfo {journal} {Phys. Rev. Lett.}}\
  }%
  \textbf{\bibinfo {volume} {104}},\ \bibinfo {pages} {196401} (\bibinfo {year}
  {2010})%
  \bibAnnoteFile{NoStop}{Nishimoto:2010}%
\bibitem{OBrien:2010}%
  \BibitemOpen
  \bibfield{author}{%
  \bibinfo {author} {\bibfnamefont{A.}~\bibnamefont{O'Brien}}, \bibinfo
  {author} {\bibfnamefont{F.}~\bibnamefont{Pollmann}},\ and\ \bibinfo {author}
  {\bibfnamefont{P.}~\bibnamefont{Fulde}},\ }%
  \bibfield{journal}{%
  \Doi{10.1103/PhysRevB.81.235115}{\bibinfo {journal} {Phys. Rev. B}}\ }%
  \textbf{\bibinfo {volume} {81}},\ \bibinfo {pages} {235115} (\bibinfo {year}
  {2010})%
  \bibAnnoteFile{NoStop}{OBrien:2010}%
\bibitem{Bergman:2006}%
  \BibitemOpen
  \bibfield{author}{%
  \bibinfo {author} {\bibfnamefont{D.~L.}\ \bibnamefont{Bergman}}, \bibinfo
  {author} {\bibfnamefont{G.~A.}\ \bibnamefont{Fiete}},\ and\ \bibinfo {author}
  {\bibfnamefont{L.}~\bibnamefont{Balents}},\ }%
  \bibfield{journal}{%
  \Doi{10.1103/PhysRevB.73.134402}{\bibinfo {journal} {Phys. Rev. B}}\ }%
  \textbf{\bibinfo {volume} {73}},\ \bibinfo {pages} {134402} (\bibinfo {year}
  {2006})%
  \bibAnnoteFile{NoStop}{Bergman:2006}%
\bibitem{Moessner:2001}%
  \BibitemOpen
  \bibfield{author}{%
  \bibinfo {author} {\bibfnamefont{R.}~\bibnamefont{Moessner}}, \bibinfo
  {author} {\bibfnamefont{S.~L.}\ \bibnamefont{Sondhi}},\ and\ \bibinfo
  {author} {\bibfnamefont{P.}~\bibnamefont{Chandra}},\ }%
  \bibfield{journal}{%
  \Doi{10.1103/PhysRevB.64.144416}{\bibinfo {journal} {Phys. Rev. B}}\ }%
  \textbf{\bibinfo {volume} {64}},\ \bibinfo {pages} {144416} (\bibinfo {year}
  {2001})%
  \bibAnnoteFile{NoStop}{Moessner:2001}%
\bibitem{Ruegg:2010}%
  \BibitemOpen
  \bibfield{author}{%
  \bibinfo {author} {\bibfnamefont{A.}~\bibnamefont{R\"uegg}}, \bibinfo
  {author} {\bibfnamefont{J.}~\bibnamefont{Wen}},\ and\ \bibinfo {author}
  {\bibfnamefont{G.~A.}\ \bibnamefont{Fiete}},\ }%
  \bibfield{journal}{%
  \Doi{10.1103/PhysRevB.81.205115}{\bibinfo {journal} {Phys. Rev. B}}\ }%
  \textbf{\bibinfo {volume} {81}},\ \bibinfo {pages} {205115} (\bibinfo {year}
  {2010})%
  \bibAnnoteFile{NoStop}{Ruegg:2010}%
\bibitem{Jose:1977}%
  \BibitemOpen
  \bibfield{author}{%
  \bibinfo {author} {\bibfnamefont{J.~V.}\ \bibnamefont{Jos\'e}}, \bibinfo
  {author} {\bibfnamefont{L.~P.}\ \bibnamefont{Kadanoff}}, \bibinfo {author}
  {\bibfnamefont{S.}~\bibnamefont{Kirkpatrick}},\ and\ \bibinfo {author}
  {\bibfnamefont{D.~R.}\ \bibnamefont{Nelson}},\ }%
  \bibfield{journal}{%
  \Doi{10.1103/PhysRevB.16.1217}{\bibinfo {journal} {Phys. Rev. B}}\ }%
  \textbf{\bibinfo {volume} {16}},\ \bibinfo {pages} {1217} (\bibinfo {year}
  {1977})%
  \bibAnnoteFile{NoStop}{Jose:1977}%
\bibitem{Hudak:1982}%
  \BibitemOpen
  \bibfield{author}{%
  \bibinfo {author} {\bibfnamefont{O.}~\bibnamefont{Hudak}},\ }%
  \bibfield{journal}{%
  \Doi{10.1016/0375-9601(82)90891-X}{\bibinfo {journal} {Physics Letters A}}\
  }%
  \textbf{\bibinfo {volume} {89}},\ \bibinfo {pages} {245 } (\bibinfo {year}
  {1982})%
  \bibAnnoteFile{NoStop}{Hudak:1982}%
\bibitem{Borisov:1985}%
  \BibitemOpen
  \bibfield{author}{%
  \bibinfo {author} {\bibfnamefont{A.~B.}\ \bibnamefont{Borisov}}, \bibinfo
  {author} {\bibfnamefont{A.~P.}\ \bibnamefont{Tankeyev}}, \bibinfo {author}
  {\bibfnamefont{A.~G.}\ \bibnamefont{Shagalov}},\ and\ \bibinfo {author}
  {\bibfnamefont{G.~V.}\ \bibnamefont{Bezmaternih}},\ }%
  \bibfield{journal}{%
  \Doi{10.1016/0375-9601(85)90791-1}{\bibinfo {journal} {Physics Letters A}}\
  }%
  \textbf{\bibinfo {volume} {111}},\ \bibinfo {pages} {15 } (\bibinfo {year}
  {1985})%
  \bibAnnoteFile{NoStop}{Borisov:1985}%
\bibitem{Gouvea:1997}%
  \BibitemOpen
  \bibfield{author}{%
  \bibinfo {author} {\bibfnamefont{M.~E.}\ \bibnamefont{Gouv\^ea}}, \bibinfo
  {author} {\bibfnamefont{G.~M.}\ \bibnamefont{Wysin}},\ and\ \bibinfo {author}
  {\bibfnamefont{A.~S.~T.}\ \bibnamefont{Pires}},\ }%
  \bibfield{journal}{%
  \Doi{10.1103/PhysRevB.55.14144}{\bibinfo {journal} {Phys. Rev. B}}\ }%
  \textbf{\bibinfo {volume} {55}},\ \bibinfo {pages} {14144} (\bibinfo {year}
  {1997})%
  \bibAnnoteFile{NoStop}{Gouvea:1997}%
\bibitem{Sinitsyn:2004}%
  \BibitemOpen
  \bibfield{author}{%
  \bibinfo {author} {\bibfnamefont{V.~E.}\ \bibnamefont{Sinitsyn}}, \bibinfo
  {author} {\bibfnamefont{I.~G.}\ \bibnamefont{Bostrem}},\ and\ \bibinfo
  {author} {\bibfnamefont{A.~S.}\ \bibnamefont{Ovchinnikov}},\ }%
  \bibfield{journal}{%
  \Doi{10.1088/0953-8984/16/20/015}{\bibinfo {journal} {Journal of Physics:
  Condensed Matter}}\ }%
  \textbf{\bibinfo {volume} {16}},\ \bibinfo {pages} {3445} (\bibinfo {year}
  {2004})%
  \bibAnnoteFile{NoStop}{Sinitsyn:2004}%
\bibitem{Louis:1999}%
  \BibitemOpen
  \bibfield{author}{%
  \bibinfo {author} {\bibfnamefont{E.}~\bibnamefont{Louis}}, \bibinfo {author}
  {\bibfnamefont{F.}~\bibnamefont{Guinea}}, \bibinfo {author}
  {\bibfnamefont{M.~P.}\ \bibnamefont{L\'opez~Sancho}},\ and\ \bibinfo {author}
  {\bibfnamefont{J.~A.}\ \bibnamefont{Verg\'es}},\ }%
  \bibfield{journal}{%
  \Doi{10.1103/PhysRevB.59.14005}{\bibinfo {journal} {Phys. Rev. B}}\ }%
  \textbf{\bibinfo {volume} {59}},\ \bibinfo {pages} {14005} (\bibinfo {year}
  {1999})%
  \bibAnnoteFile{NoStop}{Louis:1999}%
\bibitem{Runge:2007}%
  \BibitemOpen
  \bibfield{author}{%
  \bibinfo {author} {\bibfnamefont{E.}~\bibnamefont{Runge}}, \bibinfo {author}
  {\bibfnamefont{F.}~\bibnamefont{Pollmann}},\ and\ \bibinfo {author}
  {\bibfnamefont{P.}~\bibnamefont{Fulde}},\ }%
  \bibfield{journal}{%
  \Doi{10.1142/S0217979207043609}{\bibinfo {journal} {International Journal of
  Modern Physics B}}\ }%
  \textbf{\bibinfo {volume} {21}},\ \bibinfo {pages} {2215} (\bibinfo {year}
  {2007})%
  \bibAnnoteFile{NoStop}{Runge:2007}%
\bibitem{Indergand:2006b}%
  \BibitemOpen
  \bibfield{author}{%
  \bibinfo {author} {\bibfnamefont{M.}~\bibnamefont{Indergand}},\ }%
  \emph{\bibinfo {title} {Effects of strong correlations on low-dimensional and
  multi-orbital electronic systems}},\ Ph.D. thesis,\ \bibinfo {school} {ETH
  Z\"urich} (\bibinfo {year} {2006})%
  \bibAnnoteFile{NoStop}{Indergand:2006b}%
\bibitem{Seradjeh:2008}%
  \BibitemOpen
  \bibfield{author}{%
  \bibinfo {author} {\bibfnamefont{B.}~\bibnamefont{Seradjeh}}\ and\ \bibinfo
  {author} {\bibfnamefont{M.}~\bibnamefont{Franz}},\ }%
  \bibfield{journal}{%
  \Doi{10.1103/PhysRevLett.101.146401}{\bibinfo {journal} {Phys. Rev. Lett.}}\
  }%
  \textbf{\bibinfo {volume} {101}},\ \bibinfo {pages} {146401} (\bibinfo {year}
  {2008})%
  \bibAnnoteFile{NoStop}{Seradjeh:2008}%
\end{thebibliography}%

\end{document}